\documentclass[aps,pra,noshowpacs,twocolumn,preprintnumbers,superscriptaddress,floatfix]{revtex4}

\usepackage{amsmath,amssymb}
\usepackage{multirow}
\usepackage{hyperref,graphicx}
\usepackage{epsfig}
\usepackage{times}
\usepackage{pdfpages}

\begin{document}
\title{Vector Vortex Beam Emitter Embedded in a Photonic Chip}

\author{Yuan Chen}
\affiliation{Center for Integrated Quantum Information Technologies (IQIT), School of Physics and Astronomy and State Key Laboratory of Advanced Optical Communication Systems and Networks, Shanghai Jiao Tong University, Shanghai 200240, China}
\affiliation{Institute for Quantum Science and Engineering and Department of Physics, Southern University of Science and Technology, Shenzhen 518055, China}

\author{Ke-Yu Xia}  %
\thanks{keyu.xia@nju.edu.cn}
\affiliation{College of Engineering and Applied Sciences, Nanjing University, Nanjing 210093, China}
\affiliation{National Laboratory of Solid State Microstructures, Collaborative Innovation Center of Advanced Microstructures,  Nanjing University, Nanjing 210093, China}
\affiliation{Jiangsu Key Laboratory of Artificial Functional Materials, Nanjing University, Nanjing 210093, China}
\affiliation{Key Laboratory of Intelligent Optical Sensing and Manipulation (Nanjing University), Ministry of Education, China}

\author{Wei-Guan Shen}
\affiliation{Center for Integrated Quantum Information Technologies (IQIT), School of Physics and Astronomy and State Key Laboratory of Advanced Optical Communication Systems and Networks, Shanghai Jiao Tong University, Shanghai 200240, China}
\affiliation{CAS Center for Excellence and Synergetic Innovation Center in Quantum Information and Quantum Physics, University of Science and Technology of China, Hefei, Anhui 230026, China}

\author{Jun Gao}
\affiliation{Center for Integrated Quantum Information Technologies (IQIT), School of Physics and Astronomy and State Key Laboratory of Advanced Optical Communication Systems and Networks, Shanghai Jiao Tong University, Shanghai 200240, China}
\affiliation{CAS Center for Excellence and Synergetic Innovation Center in Quantum Information and Quantum Physics, University of Science and Technology of China, Hefei, Anhui 230026, China}

\author{Zeng-Quan Yan}
\affiliation{Center for Integrated Quantum Information Technologies (IQIT), School of Physics and Astronomy and State Key Laboratory of Advanced Optical Communication Systems and Networks, Shanghai Jiao Tong University, Shanghai 200240, China}
\affiliation{CAS Center for Excellence and Synergetic Innovation Center in Quantum Information and Quantum Physics, University of Science and Technology of China, Hefei, Anhui 230026, China}

\author{Zhi-Qiang Jiao}
\affiliation{Center for Integrated Quantum Information Technologies (IQIT), School of Physics and Astronomy and State Key Laboratory of Advanced Optical Communication Systems and Networks, Shanghai Jiao Tong University, Shanghai 200240, China}
\affiliation{CAS Center for Excellence and Synergetic Innovation Center in Quantum Information and Quantum Physics, University of Science and Technology of China, Hefei, Anhui 230026, China}

\author{Jian-Peng Dou}
\affiliation{Center for Integrated Quantum Information Technologies (IQIT), School of Physics and Astronomy and State Key Laboratory of Advanced Optical Communication Systems and Networks, Shanghai Jiao Tong University, Shanghai 200240, China}
\affiliation{CAS Center for Excellence and Synergetic Innovation Center in Quantum Information and Quantum Physics, University of Science and Technology of China, Hefei, Anhui 230026, China}

\author{Hao Tang}
\affiliation{Center for Integrated Quantum Information Technologies (IQIT), School of Physics and Astronomy and State Key Laboratory of Advanced Optical Communication Systems and Networks, Shanghai Jiao Tong University, Shanghai 200240, China}
\affiliation{CAS Center for Excellence and Synergetic Innovation Center in Quantum Information and Quantum Physics, University of Science and Technology of China, Hefei, Anhui 230026, China}

\author{Yan-Qing Lu} 
\thanks{yqlu@nju.edu.cn}
\affiliation{College of Engineering and Applied Sciences, Nanjing University, Nanjing 210093, China}
\affiliation{National Laboratory of Solid State Microstructures, Collaborative Innovation Center of Advanced Microstructures,  Nanjing University, Nanjing 210093, China}
\affiliation{Jiangsu Key Laboratory of Artificial Functional Materials, Nanjing University, Nanjing 210093, China}
\affiliation{Key Laboratory of Intelligent Optical Sensing and Manipulation (Nanjing University), Ministry of Education, China}

\author{Xian-Min Jin}
\thanks{xianmin.jin@sjtu.edu.cn}
\affiliation{Center for Integrated Quantum Information Technologies (IQIT), School of Physics and Astronomy and State Key Laboratory of Advanced Optical Communication Systems and Networks, Shanghai Jiao Tong University, Shanghai 200240, China}
\affiliation{CAS Center for Excellence and Synergetic Innovation Center in Quantum Information and Quantum Physics, University of Science and Technology of China, Hefei, Anhui 230026, China}

\begin{abstract}
Vector vortex beams simultaneously carrying spin and orbital angular momentum of light promise additional degrees of freedom for modern optics and emerging resources for both classical and quantum information technologies. The inherently infinite dimensions can be exploited to enhance data capacity for sustaining the unprecedented growth in big data and internet traffic, and can be encoded to build quantum computing machines in high-dimensional Hilbert space. So far much progress has been made in the emission of vector vortex beams from a chip surface into free space, however, the generation of vector vortex beams inside a photonic chip hasn't been realized yet. Here, we demonstrate the first vector vortex beam emitter embedded in a photonic chip by using femtosecond laser direct writing. We achieve a conversion of vector vortex beams with an efficiency up to 30\% and scalar vortex beams with an efficiency up to 74\% from Gaussian beams. We also present an expanded coupled-mode model for understanding the mode conversion and the influence of the imperfection in fabrication. The fashion of embedded generation makes vector vortex beams directly ready for further transmission, manipulation and emission without any additional interconnection. Together with the ability to be integrated as an array, our results may enable vector vortex beams become accessible inside a photonic chip for high-capacity communication and high-dimensional quantum information processing.
\end{abstract}

\maketitle

Light can carry both spin and orbital angular momentum (OAM). Spin angular momentum is associated with optical polarization, which is one of the most prominent and well-known properties of light. OAM is an emerging degree of freedom, who has helical wavefronts described by an azimuthal phase term $e^{ i \ell \varphi}$ \cite{Allen1992} (${\varphi}$ is the azimuthal angle in a cylindrical coordinate system). The topological charge ${\ell}$ can take any integer value represented by the number of crossed spiral wavefronts when rotating around the axis once. Due to the unlimited topological charges and the inherent orthogonality, OAM can provide larger alphabets for classical information \cite{Wang2012, Bozinovic2013, Willner2015} and quantum information processing (QIP) \cite{Barreiro2008, Dada2011, Fickler2012, Krenn2014, Mirhosseini2015, Bouchard2017, Cozzolino2018}.

Recently, there has been an increasing interest in vector vortex light beams, with a polarization varying along the azimuthal coordinate and a central optical singularity. Such beams exhibit some unique characteristics such as field structure, phase singularity and rotation invariance \cite{D'Ambrosio2012}. Vector vortex beams have been applied to wide areas including microscopy \cite{Abouraddy2006}, optical trapping \cite{Roxworthy2010}, precision measurement \cite{D'Ambrosio2013}, quantum communication \cite{D'Ambrosio2012, Vallone2014} and QIP \cite{Barreiro2010, Fickler2014, Parigi2015, D'Ambrosio2016, Cozzolino2019}. 

Large-scale applications beyond proof-of-principle demonstrations require developing integrated techniques to enable the generation \cite{Cai2012, Schulz2013, Naidoo2016, Shao2018, Liu2018}, transmission \cite{Chen2018} and even processing of vector vortex beams on a photonic chip. Several pioneering works have demonstrated on-chip generation vector vortex beams with integrated micro-ring resonators \cite{Cai2012, Schulz2013, Shao2018, Liu2018}. While the emission of vector vortex beams from the surface of integrated devices to free space has been widely investigated, the generation and transmission inside a photonic chip remain to be realized and very challenging. 

In this letter, we demonstrate a direct generation of vector vortex beams inside a photonic chip based on mode coupling. By using femtosecond laser direct writing technique \cite{Rafael2008, Szameit2010, Osellame2012}, we construct a coupled structure consisting of a single-mode waveguide and a doughnut-shaped waveguide. Gaussian beams in the single-mode waveguide evanescently couple to adjacent OAM waveguide and convert into vector vortex beams according to our coupled-mode model. In addition, we present an integrated array of such emitters and a robust generation of multiple vector vortex beams in the presence of a fluctuation of the writing laser energy.

\begin{figure}
\centering
\includegraphics[width=1.0\columnwidth]{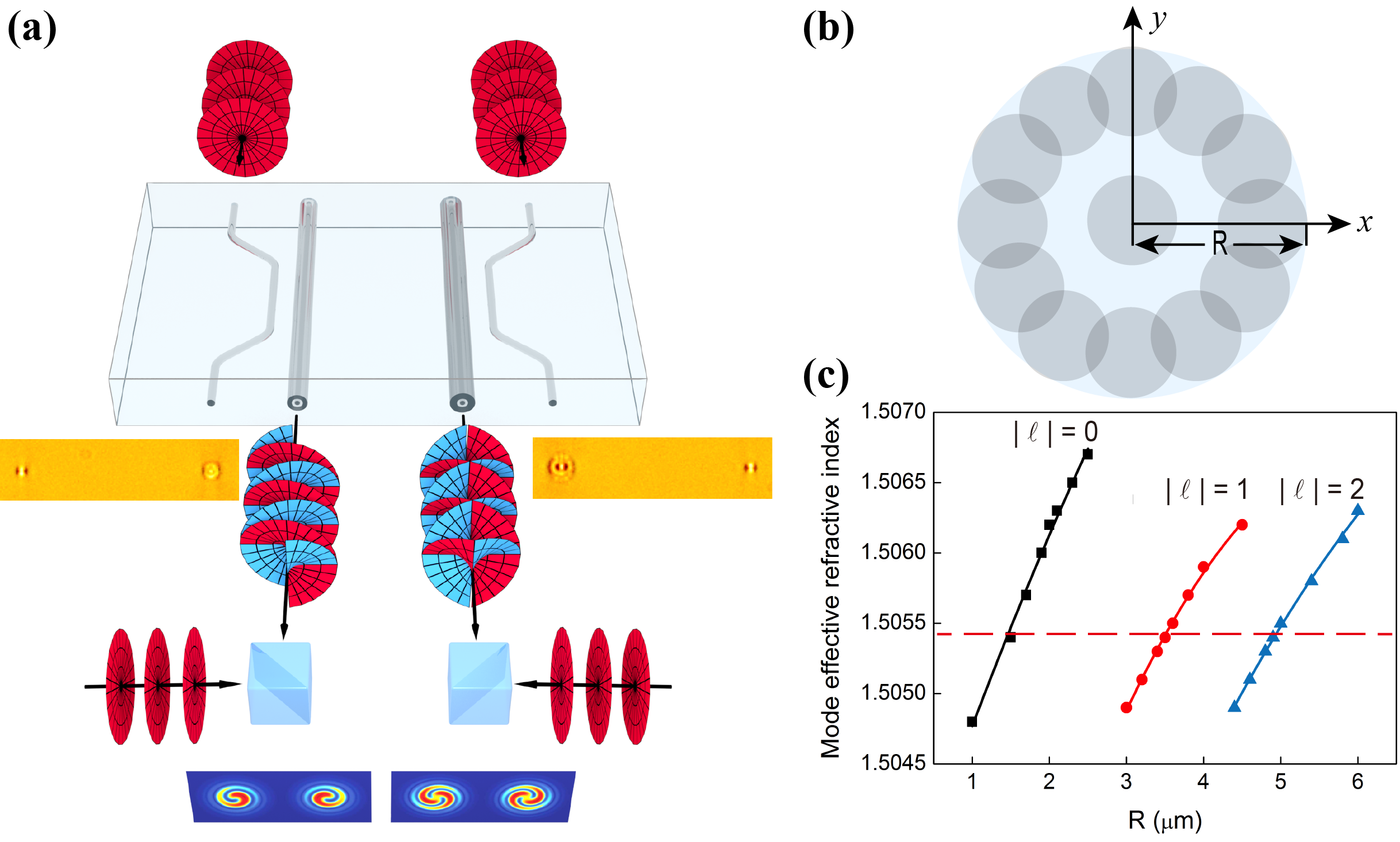}
\caption{\textbf{Experimental principle of generating vortex modes in a photonic chip.} \textbf{a.} An asymmetric direction coupler consisting of a standard single-mode waveguide and a OAM waveguide is employed to generate different order vortex beams based on phase matching. The coupling spacing and coupling length is 15 $\mu$m and 4 mm respectively. A Ti: Sapphire Oscillator centred at 780 nm is divided into two beams by inserting a beam splitter. One of the Gaussian beams is utilized as the incident light to generate vector vortex beams. The second one serves as the reference beam to measure the interference patterns with the generated vector vortex beam. A translation stage is added to tune the phase for high-contrast interference fringes and CCD is used to record the patterns. The insets show the cross-section images of a femtosecond laser-written asymmetric direction couplers. \textbf{b.} Illustration of the OAM waveguide structure. Twelve tracks constitute the annular structure with a radius R and they are overlapped to form a continuous refractive index distribution. The doughnut-shaped structure will consist of 13 waveguides when we apply an additional scan through the middle. \textbf{c.} The phase matching graph for the Gaussian modes and different order vortex modes. The black, red and blue dotted lines correspond to Gaussian modes, first-order and second-order vortex modes respectively. Horizontal red dashed line shows the phase matching for waveguides with different size to generate different order vortex beams.}
\label{Figure1}
\end{figure}

The prerequisite of generating vector vortex beams inside a photonic chip is that there exists a waveguide supporting OAM modes. The required doughnut-shaped waveguide faithfully mapping of twisted light into and out of a photonic chip has been recently achieved \cite{Chen2018}. Going beyond a straight doughnut-shaped waveguide and preservation of OAM modes, here we aim to design and construct an asymmetric direction coupler consisting of a standard single-mode waveguide and an OAM waveguide in a photonic chip. 

The generation of vector vortex fields in our experiment can be understood by expanding the standard coupled-mode theory \cite{IEEEJQE.22.988, OE.13.1515} into the vortex-mode case \cite{Suppl-III}. This model is equivalent to that in \cite{Huang1994, Okamoto2006} but is more transparent in physics. It has been widely used to study the mode coupling between optical resonators \cite{PRA.75.023814, PRA.84.013808} and waveguides \cite{OE.21.25619}.

In an ideally-made direction coupler, the single-mode (OAM) waveguide can be defined with its dielectric constant distribution $\varepsilon_s + \varepsilon_a$ ($\varepsilon_s + \varepsilon_b$), where $\varepsilon_s$ is the dielectric constant of the substrate and $\varepsilon_a$ ($\varepsilon_b$) the wanted change, symmetric around the z axis, induced by the writing laser beam. Idealy, the modes in these two waveguides have the same propagation constant $\bar\beta$. Due to the deviation of the laser energy from an axial symmetrical profile, the single-mode (OAM) waveguide has birefringence due to perturbation $\Delta\varepsilon_a$ ($\Delta\varepsilon_b$), which are tensors \cite{Suppl-III}. The envelope of field in the chip can be expanded to the superposition of eigenmodes as $\tilde{\bf E} = \sum_{m} \tilde{A}_m \tilde{\bf E}_m + \sum_n \tilde{B}_n \tilde{\bf E}_n$, where $\tilde{A}_m$ ($\tilde{B}_n$) is the amplitude of the $m$th ($n$th) eigenmode $ \tilde{\bf E}_m$ ($\tilde{\bf E}_n$) with propagation constant $\beta_m$ ($\beta_n$) in the single-mode (OAM) waveguide. $\tilde{\bf E}_m$ can be $\tilde{\bf E}_{G_{x^\prime}}$ or $\tilde{\bf E}_{G_{y^\prime}}$. $\tilde{\bf E}_n$ can be $\tilde{\bf E}_{x_{\ell}}$, $\tilde{\bf E}_{x_{-\ell}}$, $\tilde{\bf E}_{y_{\ell}}$ or $\tilde{\bf E}_{y_{-\ell}}$ with $\pm \ell \hbar$ OAM.

Starting from Maxwell's equations, we derive the general coupled-mode equations \cite{Suppl-III}
\begin{subequations} \label{eq:GCME}
  \begin{align}
    j\frac{\partial \tilde{A}_{m^\prime}}{\partial z} \!=& \Delta{\beta_{m^{\prime}}}\tilde{A}_{m^\prime}\!+\!\sum_{m \neq m^\prime} h_{m,m^\prime}  \tilde{A}_m\!+\!\sum_n \kappa_{n,m^\prime} \tilde{B}_n \;,\\
    j \frac{\partial \tilde{B}_{n^\prime}}{\partial z} \!=& \Delta{\beta_{n^{\prime}}}\tilde{B}_{n^\prime}\!+\! \sum_{n \neq n^\prime} h_{n,n^\prime} \tilde{B}_n \!+\! \sum_m \kappa_{m,n^\prime} \tilde{A}_m\;,
  \end{align}
\end{subequations}
where $\Delta{\beta_{m^{\prime}}}=\beta_{m^{\prime}}-\bar{\beta}+ \delta_{m^{\prime}}$ and $\Delta{\beta_{n^{\prime}}}=\beta_{n^{\prime}}-\bar{\beta}+ \delta_{n^{\prime}}$. Ideally, $\beta_{m^{\prime}}=\bar{\beta}$ and $\beta_{n^{\prime}}=\bar{\beta}$. Here, we neglect the Butt couplings because they are orders smaller than the shifts and other couplings.
The propagation constant shifts $\delta_{m^\prime}$ ($\delta_{n^\prime}$), the couplings $ h_{m,m^\prime}$ ($h_{n,n^\prime}$) between the eigenmodes in the single-mode (OAM) waveguide, and the couplings $\kappa_{n,m^\prime}$ ($\kappa_{m,n^\prime}$) between the waveguides are given by
\begin{subequations}
\begin{align}
 \delta_{X} \!=\! & \frac{k_0^2}{2 \beta_X N_{X}} \int \tilde{\mathbf{E}}_X^* \cdot (\varepsilon_p \!+\! \Delta\varepsilon_b \!+\! \Delta\varepsilon_a) \cdot \tilde{\mathbf{E}}_X dx dy \;, \\
 h_{Y,X} \!=\! &  \frac{k_0^2}{2 \beta_X N_X} \int \tilde{\mathbf{E}}_X^* \cdot (\varepsilon_p \!+\! \Delta\varepsilon_b \!+\! \Delta\varepsilon_a) \cdot \tilde{\mathbf{E}}_Y dx dy  \;, 
\end{align}
\end{subequations}
where $\int \tilde{\mathbf E}_X \cdot \tilde{\mathbf E}^*_X dx dy = N_X$ and $\{X,Y,p\}=\{m^\prime,m,b\}$ ($\{X,Y,p\}=\{n^\prime,n,a\}$) for the single-mode (OAM) waveguide, and
\begin{equation}
 \kappa_{Y,X} \!=\frac{k_0^2}{2 \beta_X N_X} \int \tilde{\mathbf{E}}_X^* \cdot (\varepsilon_p \!+\! \Delta\varepsilon_a \!+\! \Delta\varepsilon_b) \cdot \tilde{\mathbf{E}}_Y dx dy  \;, 
\end{equation}
where $\{X,Y,p\}=\{m^\prime,n,a\}$ ($\{X,Y,p\}=\{n^\prime,m,b\}$), $k_0$ is the propagation constant of light in free space. Obviously, the birefringence resulting from $\Delta\varepsilon_a$ and $\Delta\varepsilon_b$ changes the propagation constant shifts and all coupling coefficients.

Gaussian beams in the single-mode waveguide can be evanescently coupled to the adjacent doughnut-shaped waveguide and converted into vector vortex beams according to our coupled-mode theory (see Fig. 1a). However, the complete modal conversion can only happen under $\Delta\beta_{m^{\prime}, n^{\prime}}\!=\! \beta_{m^{\prime}}-\beta_{n^{\prime}}\!=\!0$ \cite{Ismaeel2014}. Note that $\beta_{m^{\prime}}$ ($\beta_{n^{\prime}}$) is mode dependent. Therefore, the phase matching condition should be subject to exciting the specific order of vector vortex beams. Instead of propagation constants $\beta_{m^{\prime}}$ and $\beta_{n^{\prime}}$, we choose more convenient effective refractive index $\it n_{eff}$=${\beta}/{k}$. In our experiment, we fix the size of single-mode waveguide and tune the effective refractive index via changing the size of OAM waveguide \cite{Ismaeel2014, Luo2014, Wang2014, Yang2014, Mohanty2017}. The OAM waveguide structure with radius R is shown in Fig. 1b. 

\begin{figure}
\centering
\includegraphics[width=0.99\columnwidth]{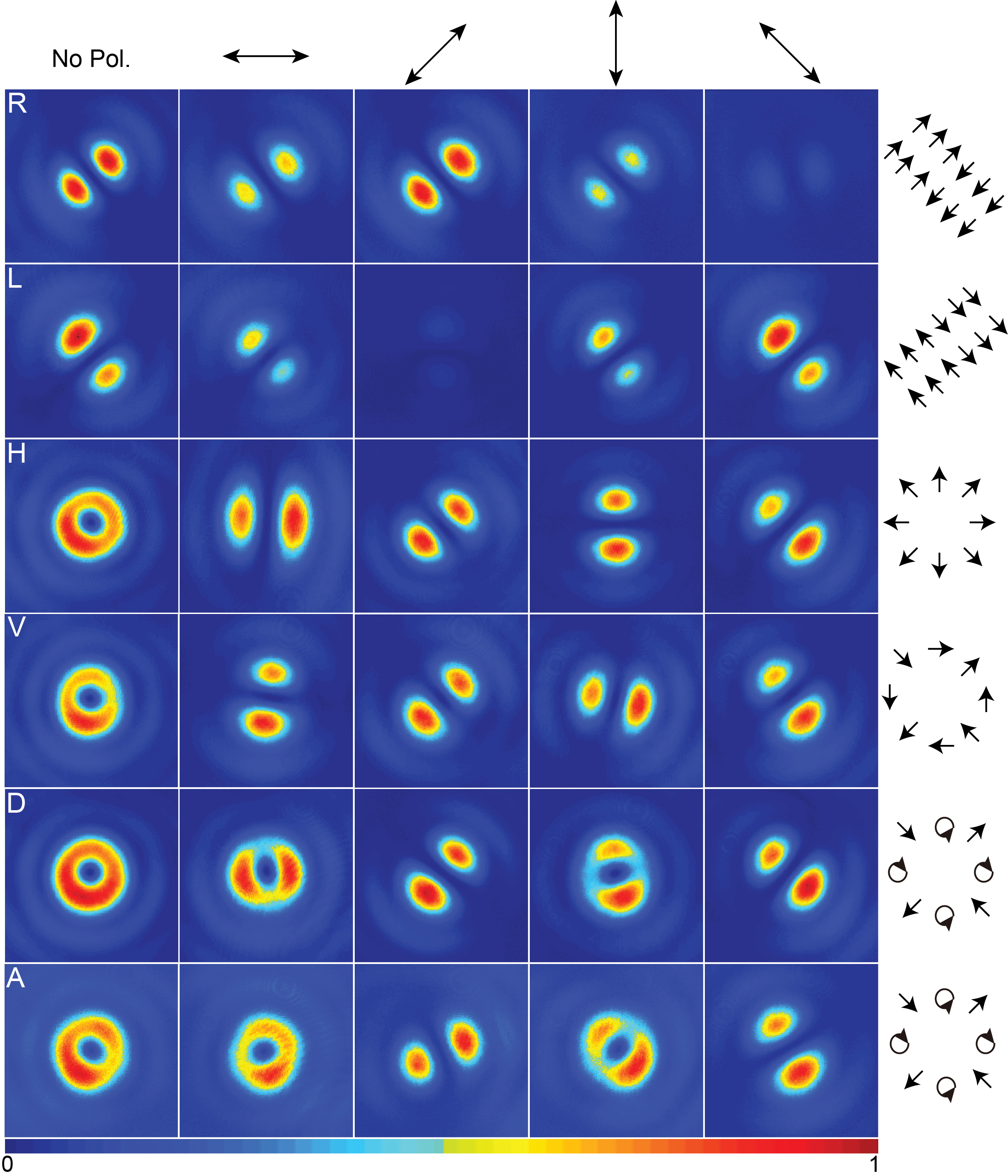}
\caption{\textbf{Experimental results of generating first-order vortex beams.} Measured intensity distribution for right circularly (left circularly, horizontally, vertically, diagonally, anti-diagonally)-polarized Gaussian beams and the corresponding intensity distributions after polarization analysis are shown in the first (second, third, fourth, fifth, sixth) row. The direction of polarization projection is indicated by the arrows and the spatial polarization distribution is shown at the end of each row. The resulting modes have different spatial polarization distribution for different input polarized Gaussian beams, which suggests there exists birefringence in the doughnut-shaped waveguide.}
\label{Figure2}
\end{figure}

To find the optimal radius at which Gaussian modes are phase matched with vortex modes, we calculate the effective refractive indices for different order of vortex modes \cite{Ismaeel2014, Mohanty2017} (see Fig. 1c). The phase matching graph indicates that OAM waveguide should be written with a radius of around 3.5 $\mu$m (4.9 $\mu$m), which can match Gaussian modes to generate first-order (second-order) vortex modes. The calculated results are instructive, however, since the fabrication and coupling of two different types of waveguides are involved and laser-matter interaction is quite complicated with many involved physical processes \cite{Szameit2010, Osellame2012}, a detailed scan of the size of OAM waveguide near the estimated value is essential to achieve an efficient mode conversion. Finally, we obtained high-quality first-order and second-order vortex modes at R $\simeq$ 3.7 $\mu$m and R $\simeq$ 5.0 $\mu$m, respectively. %

To analyze the evolution of field in the OAM waveguide, we divide the asymmetric direction coupler into three segments: the input ($0\le z\le L_{1}$), coupling ($L_{1}\le z \le L_{1}+L_{cp}$) and output ($L_{1}+L_{cp}\le z \le L$) regions, where $L$ is the length of chip. In the input and output regions, the two waveguides are uncoupled. However, due to the non-zero $\Delta\varepsilon_a$ and $\Delta\varepsilon_b$, the field changes during propagating in these regions. The modes in two waveguides interact with each other in the coupling region. The evolution of field can be described by Eqs.~1. 

The incident Gaussian field can be written as $
{\bf E}_{in}(0)\!=\! \tilde{A}_{G_{x^\prime}}(0)\tilde{\bf E}_{G_{x^\prime}}\!+\!\tilde{A}_{G_{y^\prime}}(0)\tilde{\bf E}_{G_{y^\prime}}$. The field can be determined by the input and the evolution matrix $M$ \cite{Suppl-IV}, which is dependent on the structure of the waveguides and is a function of $\Delta{\beta_{m^{\prime}}}$, $\Delta{\beta_{n^{\prime}}}$, $ h_{m,m^\prime}$, $h_{n,n^\prime}$, $\kappa_{n,m^\prime}$ and $\kappa_{m,n^\prime}$ \cite{Suppl-III}. The output field can also be expanded as the superposition of eigenmodes in the OAM waveguide as
$\tilde{\bf E}(L) \!=\! \gamma_{x_{\ell}}\tilde{\bf E}_{x_{\ell}}\!+\gamma_{x_{-\ell}}\tilde{\bf E}_{x_{-\ell}}\!+\gamma_{y_{\ell}}\tilde{\bf E}_{y_{\ell}}\!+\gamma_{y_{-\ell}}\tilde{\bf E}_{y_{-\ell}}$,  with 
\begin{equation}\label{eq:GaussianL3}
\begin{bmatrix}\gamma_{x_{\ell}}\\ \gamma_{x_{\!-\!\ell}} \\ \gamma_{y_{\ell}} \\ \gamma_{y_{\!-\!\ell}} \end{bmatrix}^{T}\!=\!\begin{bmatrix}\tilde{A}_{G_{x^\prime}}(0)\\ \tilde{A}_{G_{y^\prime}}(0) \\ 0 \\ 0 \end{bmatrix}^{T} M_1^{T}(L_{1})M^{T}_{cp}(L_{cp})M_2^{T}(L_{2}) \;,
\end{equation} 
where $\tilde{\mathbf{E}}_{x_{\pm \ell}}={E}_{x_{\pm \ell}}e^{\pm j\ell\phi}{\mathbf e}_{x}$, $\tilde{\mathbf{E}}_{y_{\pm \ell}}={E}_{y_{\pm \ell}}e^{\pm j\ell\phi}{\mathbf e}_{y}$, $M_1$ ($M_2$, $M_{cp}$) describes the evolution of field in the input (output, coupling) region, $T$ means the transpose of a matrix.

When right circularly-polarized Gaussian beams incident and evanescently couple into the OAM waveguide, a two-lobe intensity distribution is obtained (see Fig.~2). After projecting to horizontal and vertical polarization, the two-lobe intensity distribution in near-diagonal direction suggests that $\gamma_{x_{-\ell}}{E}_{x_{-\ell}}  \simeq j\gamma_{x_{\ell}}{E}_{x_{\ell}}$ and $\gamma_{y_{-\ell}}{E}_{y_{-\ell}}  \simeq  j\gamma_{y_{\ell}}{E}_{y_{\ell}}$. Note that the intensity after anti-diagonal polarization projection is relatively small comparing with diagonal polarization with an extinction ratio up to 10.2 dB observed in the experiment. Obviously, this is a scalar light field with diagonal polarization. The small component on the anti-diagonal polarization means $\tilde{\bf E}(L) \cdot {\mathbf e}_{a}\simeq0$, so that $\gamma_{y_{\ell}}E_{y_{\ell}}\simeq\gamma_{x_{\ell}} E_{x_{\ell}}$ \cite{Suppl-IV}. 

With an input of left circularly-polarized Gaussian beam, the generated vortex beam also exhibits a scalar light field with anti-diagonal polarization. Although the intensity distribution is uneven, the two lobes near the anti-diagonal direction after respectively projecting to horizontal, vertical polarization indicate that $\gamma_{x_{-\ell}}{E}_{x_{-\ell}} \simeq -j\gamma_{x_{\ell}}{E}_{x_{\ell}}$ and $\gamma_{y_{-\ell}}{E}_{y_{-\ell}} \simeq -j\gamma_{y_{\ell}}{E}_{y_{\ell}}$. The intensity after diagonal polarization projection is relatively small comparing with anti-diagonal polarization with an extinction ratio up to 9.6 dB. 
The component with diagonal polarization is $\tilde{\bf E}(L) \cdot {\mathbf e}_{d}\simeq0$, meaning $\gamma_{y_{\ell}}{E}_{y_{\ell}}\simeq -\gamma_{x_{\ell}}{E}_{x_{\ell}}$ \cite{Suppl-IV}. 

When the asymmetric direction coupler excited by horizontally-polarized Gaussian beams, we obtain good circularly symmetric first-order vector vortex modes, whose intensity distribution are annular with a dark core in the center. Projecting the generated beam to different polarization with a polarizer, the two-lobe pattern is formed and rotates with the polarizer, which implies that it is a cylindrical vector vortex beam with radial polarization \cite{Dorn2003, Maurer2007, Naidoo2016}. This experimental result manifests that $\gamma_{x_{-\ell}}{E}_{x_{-\ell}} \simeq \gamma_{x_{\ell}}{E}_{x_{\ell}}$, $\gamma_{y_{-\ell}}{E}_{y_{-\ell}} \simeq -\gamma_{y_{\ell}}{E}_{y_{\ell}}$ and $\gamma_{y_{\ell}}{E}_{y_{\ell}} \simeq -j\gamma_{x_{\ell}}{E}_{x_{\ell}}$ \cite{Suppl-IV}.

\begin{figure}[htb!]
\centering
\includegraphics[width=1.0\columnwidth]{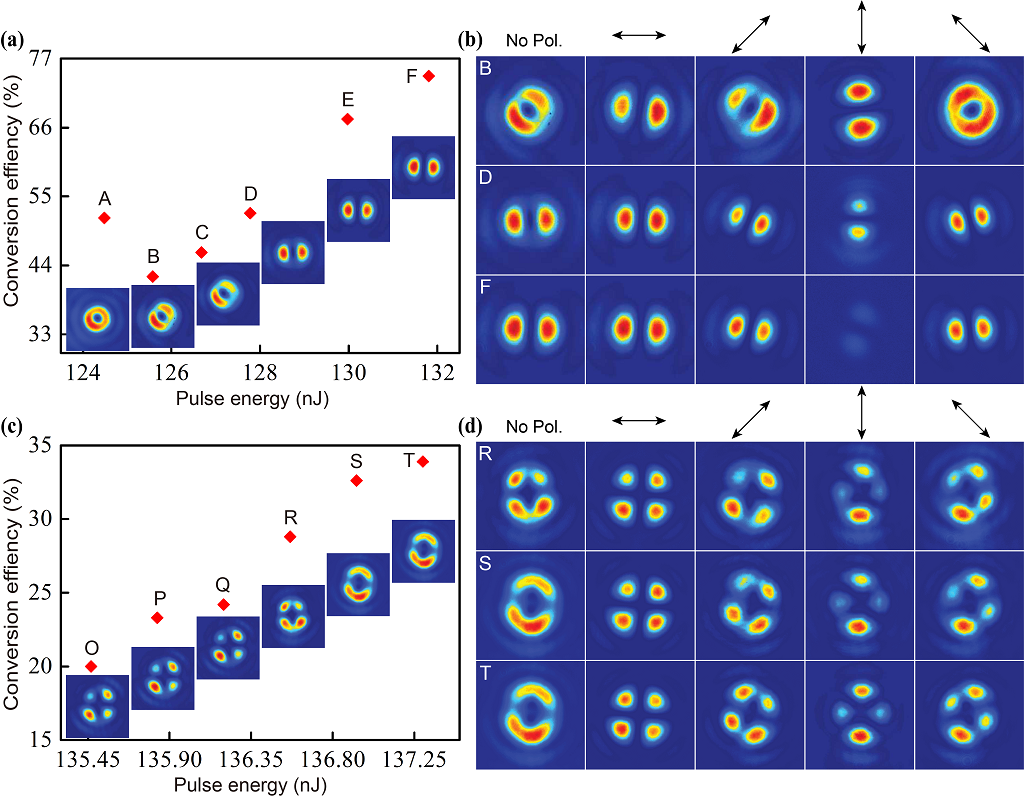}
\caption{\textbf{The evolution of resulting vortex modes when increasing the writting pulse energy.} \textbf{a.} Conversion effiency and first-order vortex modes versus the writting pulse energy. \textbf{b.} The polarization analysis on the typical states B, D and F. It is obvious that, when we increase the pulse energy, the resulting beam changes from vector vortex to scalar vortex. \textbf{c.} Conversion effiency and second-order vortex modes versus the writting pulse energy. \textbf{d.} The polarization analysis on the typical states R, S and T. The direction of polarization projection is indicated by the arrows.}
\label{Figure3}
\end{figure}

Similarly, we generate a ${\pi}$-vector vortex beam \cite{Roxworthy2010, D'Ambrosio2016} by coupling a vertically-polarized Gaussian beam into the doughnut-shaped waveguide. Our polarization analysis \cite{Cardano2012, Maurer2007} indicates that the vector vortex mode is linearly polarized at each local position, but with a local polarization direction changing between radial and azimuthal direction (see Fig. 2). We also find that when injecting a diagonally (anti-diagonally)-polarized Gaussian beam, the generated mode shows an annular intensity distribution when projected to horizontal and vertical polarization, while it shows a two-lobe intensity distribution when projected to a diagonal and anti-diagonal polarization. The resulting modes have different spatial polarization distribution for Gaussian beams with different input polarization, which suggests there exists birefringence in the doughnut-shaped waveguide.

\begin{figure}
\centering
\includegraphics[width=0.97\columnwidth]{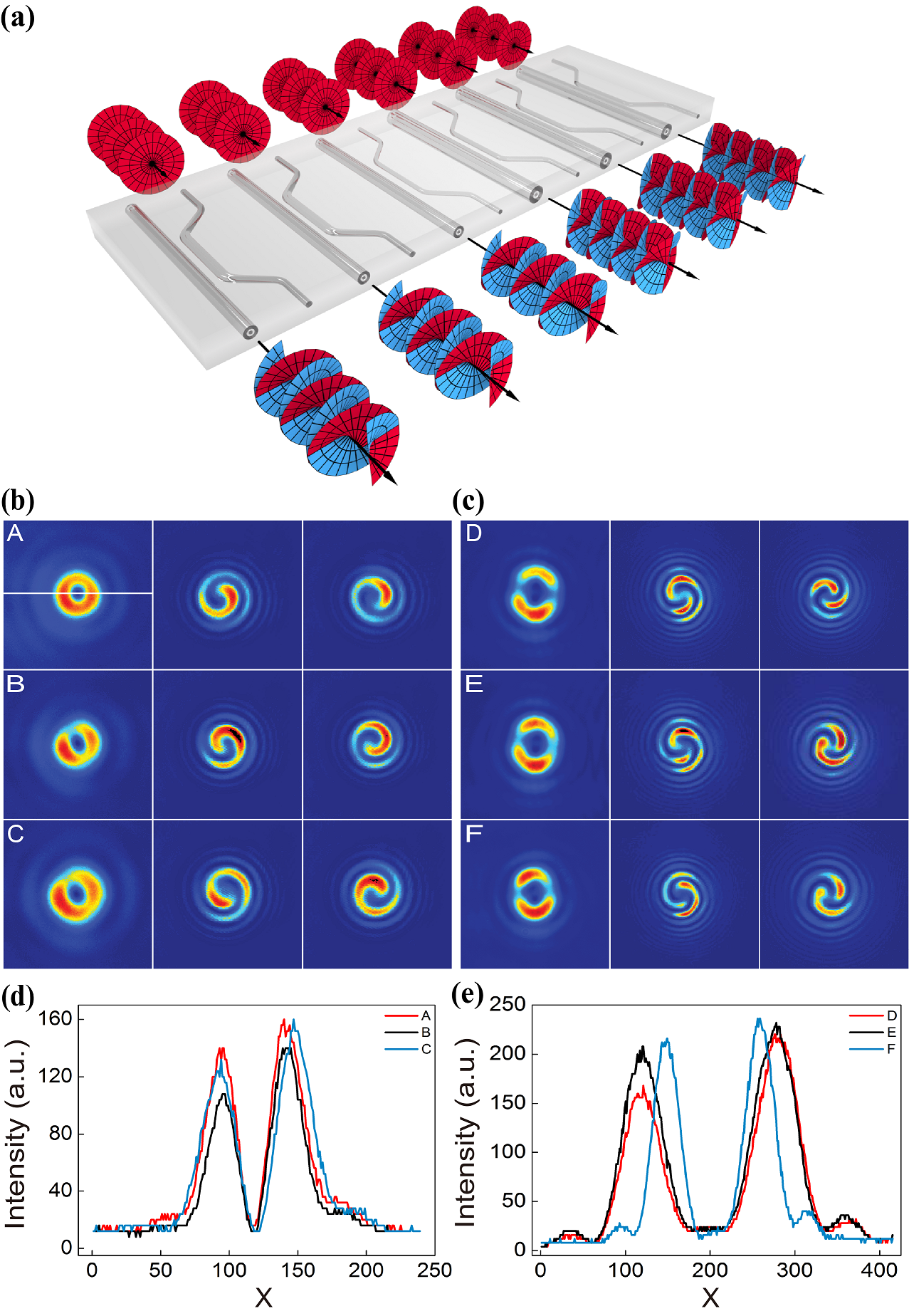}
\caption{\textbf{Experimental results of vortex beam emitter array.} \textbf{a.} Illustration of an array consisting of three asymmetric directional couplers for first- and second-order vector vortex beam emitters respectively. \textbf{b.} The first column: intensity patterns generated from the first-order vortex beam array. The difference in their shape can be attributed to slight differences in the writting pulse energy and fabrication variations. The clockwise (counterclockwise) spiral interference patterns with one-arm are shown in the second (third) column for first-order vector vortex beams. \textbf{c.} The first column: intensity patterns generated from the second-order vortex beam array. The clockwise (counterclockwise) spiral interference patterns with two-arm are shown in the second (third) column. Even with an artificially introduced energy fluctuation of the write beam, about 1.465 nJ (0.4 nJ) in generating first-order (second-order) vector vortex beams, we still can obtain almost identical emission and a robust generation of multiple vector vortex beams against the fluctuation of the writing laser energy. \textbf{d.} Radial intensity distribution extracted from the first-order vortex beam along the white radial direction corresponds to A, B and C. \textbf{e.} Radial intensity distribution extracted from the second-order vortex beam D, E and F.}
\label{Figure4}
\end{figure}

A variation of the coupling between the waveguides by changing the energy of the writing laser have been observed in Fig. 3, which can be reflected into a change in propagation constant. Our thoery, in good agreement with previous experiments \cite{Diener2018, Tang2019}, shows that this change is linearly dependent on the energy variation of the writing laser \cite{Suppl-I}.

When we increase the pulse energy, the distribution of optical axis has no substantial change \cite{Suppl-II}, and the conversion efficiency of first-order vortex modes increases from 42.2\% to 74.2\% (see Fig. 3a). Meanwhile, the resulting modes become pure scalar vortex modes. We can see that the generated modes are all good vector vortex modes in a range of writing pulse energy. We further perform comprehensive polarization analysis on the typical states (B, D and F) generated with different pulse energy (see Fig. 3b). We can observe a clear trend to single-polarization vortex beams and a high extinction ratio at high writing pulse energy (F). 

By increasing the radius of the doughnut-shaped waveguide, we can tune the phase matching condition for second-order vortex modes (see Fig. 3c). We find that there also exists a sweet spot of the writing pulse energy, where second-order vector vortex beams can be well observed and the conversion efficiency can be optimized before apparently becoming a scalar vortex beam. Before the efficiency reaches 33.9\%, the polarization analysis on R, S and T reveals no distinct changes of polarization distribution (see Fig. 3d). 

To demonstrate the potential towards large-scale integration, we construct an array consisting of three such asymmetric directional couplers as first-order and second-order vector vortex beam emitters respectively (see Fig. 4a). Owing to the complex physical process of femtosecond laser micromachining, many parameters have been found strongly affecting the resulting morphology and therefore the characteristics of waveguides \cite{Osellame2012}. To keep all the waveguide identical, we lock all the parameters of the laser, the translation stages and the environment. We show that, even with an artificially introduced fluctuation, about 1.465 nJ (0.4 nJ) in generating first-order (second-order) vector vortex beams, we still can obtain almost identical emission (see Fig. 4b and 4c). The expected clockwise or counterclockwise spiral interference patterns clearly identify the ingredient of three output states in first- and second-order vector vortex modes. We extract the radial intensity distribution from the first- and second-order modes along the white radial direction (see Fig. 4d and 4e). The relatively balanced intensity distribution once again implies a robust generation of vortex beams against the fluctuation of the writing laser energy.

In summary, we demonstrated the first vector vortex beam emitter embedded in a photonic chip by using femtosecond laser direct writing. By engineering the phase matching condition, we generated the first- and second-order vector vortex beams, and achieved very high conversion efficiency by tuning the writing pulse energy. The fashion of embedded generation makes vector vortex beams directly ready for further transmission, manipulation and emission without any additional interconnection. We also demonstrated an integrated array of such emitters and a robust generation of multiple vector vortex beams, which is crucial towards high-capacity communication and high-dimensional \cite{Mehul2016, Erhard2018} QIP.

This emerging field of integrated photonics of vector vortex beams has many open problems to be addressed. The evanescent light coupling or splitting between two vector vortex waveguides is a primary goal, which may facilitate the design and fabrication of many novel integrated devices of vector vortex beams, especially may enable on-chip quantum interference of vector vortex beams. Quantum inference between the transverse spatial modes has been observed in multimode waveguide \cite{Mohanty2017}, opening possibility to realize on-chip Hong-Ou-Mandel interference for vector vortex modes \cite{D'Ambrosio2017}. The efficient generation of pure scalar vortex modes in our experiment promises the potential for preparing indistinguishable vortex photonic states and observing the interference with high visibility. In addition, introducing chiral structure to the OAM waveguide may allow controllably generate certain chiral vortex states.

\section*{Acknowledgements} The authors thank Miles Padgett, Nathan Langford, Si-Yuan Yu and Jian-Wei Pan for helpful discussions. This work was supported by National Key R\&D Program of China (2019YFA0308700, 2017YFA0303700); National Natural Science Foundation of China (NSFC) (61734005, 11761141014, 11690033, 11890704, 11874212); Science and Technology Commission of Shanghai Municipality (STCSM) (17JC1400403); Shanghai Municipal Education Commission (SMEC)(2017-01-07-00-02-E00049); X.-M.J. acknowledges additional support from a Shanghai talent program.

\clearpage
\newpage

\bigskip

\onecolumngrid
\section*{\large Supplemental Material: Vector Vortex Beam Emitter Embedded in a Photonic Chip}
\setcounter{figure}{0}
\setcounter{table}{0}
\setcounter{equation}{0}
\renewcommand{\tablename}{Supplementary Table}

\renewcommand{\thetable}{\arabic{table}}
\renewcommand{\theequation}{{S}\arabic{equation}}

The whole Supplemental Material consists of four parts. In the first part \textbf{I}, we give the relationship between the writing laser energy and the propagation constant. The optical axis distribution of OAM waveguide is shown in the second part \textbf{II}. In the third part \textbf{III}, we derive the general coupled-mode equations including six eigenmodes and also present a formula to calculate all involved parameters. Based on this coupled-mode theory, the evolution of field in the OAM waveguide can be derived in the fourth part \textbf{IV}. 
 
\section{\textbf{The relationship between the writing laser energy and the propagation constant}} 
A variation of the coupling between the waveguides by changing the speed of writing a waveguide and the pulse energy of the writing laser is observed. The reason is that the change of the writing laser energy causes the change of the effective refractive index of the written waveguide and therefore the propagation constant. Such behavior can be reflected into a change in the coupled parameters. The writing laser energy, which causes heating effect in material and melts it, deposited in the glass is obviously dependent on the totally accumulated laser pulse energy $W$. Thus, the melting degree of material can be controlled by the effective pulse number per unit time in the laser spot size \cite{OsellameR2012, RajeshR2010}. 

It has been found that changing the writing speed $v$ controls the amount of deposited laser energy through the number of pulses focused at the same spot in the glass and therefore directly varies the mode propagation constant \cite{LebugleR2015}. The difference of propagation constants has also been reported as a function of the writing speed difference $ v_2 - v_1$ of waveguide 2 with respect to waveguide 1 \cite{DienerR2018}, where $v_2$ and $v_1$ are the speeds of writing the waveguide 2 and the waveguide 1, respectively. Tang and co-workers \cite{TangR2019} present a linear relationship between the change of propagation constant $\delta \beta$ of waveguide and the writing speed $\delta v$ as $\delta \beta=a (v_2-v_1)+b$ (a, b are fitting constants).  

Based on the reported experiments and ours, here we present a simple model to derive the relationship of the propagation constant and the pulse energy and the writing speed $v$. We start our model from intuitive but reasonable assumption: the change of the relative dielectric constant $\delta \varepsilon ({\bf r})$ of material at the transverse position ${\bf r}$ of the waveguide cross section is much less than unity, and thus proportional to the total laser energy $W({\bf r})$ absorbed by a small volume $\delta V({\bf r})$ at ${\bf r}$ that $\delta \varepsilon ({\bf r})  = \eta W({\bf r}) / \delta V({\bf r}) \varepsilon_0$ with the efficiency $\eta$ of laser energy converting to heat and material strain, typically determined by the material, and the vacuum permittivity $\varepsilon_0$. Obviously, the dielectric constant change at ${\bf r}$ is actually proportional to the energy density $W_D({\bf r}) = W({\bf r}) / \delta V({\bf r})$ at the position. The waveguide is written in a material with a refractive index $n_s \gg \sqrt{\delta \varepsilon ({\bf r})}$. The refraction index at ${\bf r}$ is given by
\begin{equation}\label{eq:nr}
n({\bf r}) = \sqrt{n_s^2 + \delta \varepsilon ({\bf r}) } \approx n_s +  \delta \varepsilon ({\bf r})  /2n_s \;.
\end{equation}

In this sense, the shape and the refractive profile of the waveguide are dependent on the transverse intensity profile of the laser pulse. We consider the average change of the relative dielectric constant of material over the laser writing  region $V$. This region has an effective volume $V_\text{eff}$. Then, we get 
\begin{equation}
\langle \delta\varepsilon \rangle = \int_V \delta\varepsilon({\bf r}) d{\bf r}/V_\text{eff} = \eta W_D/ \varepsilon_0 \;,
\end{equation}
where $W_D$ is the average energy density over the waveguide.
 According to the theory for a weak-guiding waveguide, the propagation constant of the waveguide is 
\begin{equation} \label{eq:Beta}
\beta = k_0 n_\text{eff} \approx k_0^\prime \left(n_s +  \langle \delta\varepsilon \rangle  /2n_s \right) \;,
\end{equation}
where $k_0$ is the propagation constant of laser in free space, and $k_0^\prime$ is a modified propagation constant, approximately being $k_0$ for a weak guiding waveguide in our experiment. This modification is due to the simple linear approximation form above. 

Next, we find out the laser energy density $W_D({\bf r})$ during a waveguide writing  period $\tau_w$. For simplicity, we consider all laser pulses identical. We assume that laser pulses have the pulse duration $\tau_p$, the waist  $w_0$, and the intensity distribution $I({\bf r},t)$, which is the spatial laser energy density at position ${\bf r}$ at time $t$.  The energy included in a single pulse is 
\begin{equation}\label{eq:PulseE}
E_\text{sp} = \int I({\bf r}, t) d{{\bf r}} dt \;.
\end{equation}
We consider that the profile of the laser pulse matches the wanted waveguide profile. During a writing period, we make a $L$-long waveguide  with a waist $w_0$ by applying $N$ pulses. Then, the average energy density has the form
\begin{equation} \label{eq:EDensity}
\begin{split}
W_D & = \frac{N E_\text{sp}}{w_0 L} \\
         & =  \frac{E_\text{sp} \tau_w f_\text{rp}}{w_0 L} \\
         & =  \frac{E_\text{sp} f_\text{rp}}{w_0 v} \;,
\end{split}
\end{equation}
where $f_\text{rp}$ is the repeating rate of the laser pulses.

Substituting Eq.~\ref{eq:EDensity} into Eq.~\ref{eq:Beta}, we have the propagation constant of the written waveguide, given by
\begin{equation}
\beta = k_0^\prime \left( n_s + \frac{\eta E_\text{sp} f_\text{rp}}{2 \varepsilon_0 n_s w_0 v }  \right) \;.
\end{equation}
This propagation constant normally is the sum of the wanted value in the target design and some variation. It is clear now that the variation of the pulse energy and the writing speed can cause the change of the propagation constant in the way as
\begin{equation}\label{eq:DBeta}
\delta\beta = \frac{k_0^\prime \eta f_\text{rp}}{2 \varepsilon_0 n_s w_0 v} \delta E_\text{sp} - \frac{k_0^\prime\eta E_\text{sp} f_\text{rp}}{2\varepsilon_0 n_s w_0 v^2} \delta v + \xi \;,
\end{equation}
where $\xi$ can be due to some unknown fluctuation or the rest strain in the material. Here, the change of the propagation constant $\delta \beta$ can be caused by the changes of the laser pulse energy or the difference of the writing speed of two waveguides by different control parameters. Our theoretical result is in agreement with the existing experiments \cite{DienerR2018, TangR2019}. According to our analytical formula, if the intensity profile of the laser pulses is different along two orthogonal directions, say $x$ and $y$ axes, then the written waveguide will have two different propagation constants for modes polarized along $x$ and $y$, possessing birefringence. 

Assuming that we write a target structure with a single-pulse laser energy $E_\text{sp}^{(T)}$, the change of the dielectric constant in the waveguide region is $\varepsilon_w$. If the writing laser pulse energy has a fluctuation $\delta E_\text{sp}$, then this fluctuation will cause a small variation $\delta \varepsilon$ proportional to $\delta E_\text{sp}$. In this, the dielectric constant of the waveguide becomes $\varepsilon_w + \delta\varepsilon$.

{\color{black}\section{\textbf{The optical axis distribution}} 

\renewcommand{\thefigure}{S\arabic{figure}}
\begin{figure*}[htb!]
\centering
\includegraphics[width=0.77\columnwidth]{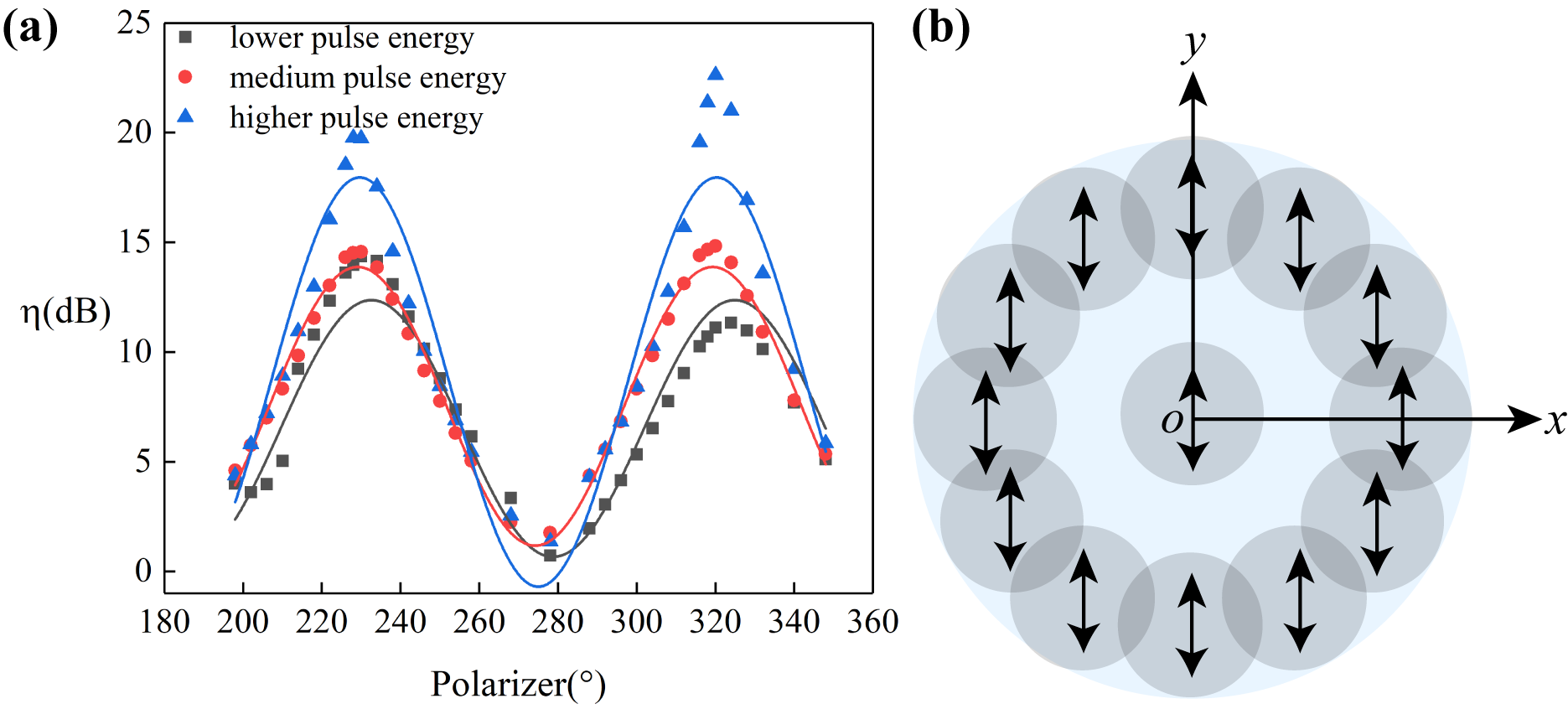}
\caption{\textbf{The optical axis distribution}. \textbf{a.} The extinction ratio changes with incident polarization. In the scenarios of low, medium and high pulse energy, we measure the extinction ratio dependence on the incident polarization. The black, red and blue dots are experimental data, and the black, red and blue lines are the curves fitted with a sine function, respectively. \textbf{b.} The optical axis distribution of each single-mode track in doughnut-shaped waveguide.}
\label{Figure S1}
\end{figure*}

Since the doughnut-shaped waveguide supports single-mode faithful transmission \cite{ChenY2018}, we inject different polarization into the waveguide before the chip, and perform polarization analysis for the output single-mode. If the optical axis distribution is along the radial direction, then we can get a high extinction ratio in each direction. In fact, we only achieve a high extinction ratio in two perpendicular directions as shown in Fig.~\ref{Figure S1}. Thus, the optical axis of each single-mode track constituting the doughnut-shaped waveguide should be fixed along the vertical or horizontal direction shown in Fig.~\ref{Figure S1}. In addition to the small difference in spatial position of each single-mode track, their direct-written parameters are exactly the same. Therefore, their optical axis orientations should be the same in theory instead of distributing along the radial direction. In addition, we need to pay attention to the fact that the change of pulse energy does not affect the optical axis distribution, but only changes the composition of the output state.}

It can be clearly seen from Fig.~\ref{Figure S1} that the orthogonal component can be excited when an electric field ${\bf E}_\text{in}$ incident on a waveguide with birefringence does not follow the optical axis of the waveguide. Generally, we consider that a birefringent waveguide with optical axes along $x$ and $y$. An input field ${\bf E}_\text{in}$ is polarized along $a_x \hat{e}_x +b_y \hat{e}_y$ with $a_x b_y \neq 0$ propagating in this waveguide, where $\hat{e}_x$ and  $\hat{e}_y$ are the unit vector along the directions $x$ and $y$, respectively. The orthogonally polarized but equal amplitude field is ${\bf E}_o = b_y \hat{e}_x - a_x \hat{e}_y$ and ${\bf E}_\text{in} \cdot {\bf E}_o = 0$. After propagating a distance $L$, the components along $x$ and $y$ accumulates different phases $\phi_x$ and $\phi_y$ that the electric field becomes ${\bf E}_L = a_x e^{i\phi_x}\hat{e}_x + b_y e^{i\phi_y} \hat{e}_y$. The projection of the field ${\bf E}_L$ to the orthogonal direction of ${\bf E}_\text{in}$ is $( b_y \hat{e}_x - a_x \hat{e}_y) \cdot \left( a_x e^{i\phi_x}\hat{e}_x + b_y e^{i\phi_y}\hat{e}_y\right) = a_x b_y \left(e^{i\phi_x} - e^{i\phi_y} \right)$, which is nonzero when $\phi_x \neq \phi_y + 2 q \pi$ with an integer $q$ for the birefringent waveguide. This means that the two orthogonal modes will couple to each other in a birefringent medium when either of them has nonzero components in the optical axes of medium.

\section{\textbf{General coupled-mode equations between two waveguides}}
In this section, we present a coupled-mode theory modeling the coupling between a Gaussian mode in the single-mode waveguide and the vortex mode possessing orbital angular momentum in the donut-shaped waveguide. By using femtosecond laser direct writing technique, we fabricate a single-mode waveguide and an OAM-mode waveguide, schematically depicted in Fig.~\ref{Figure S2}. Then, we experimentally demonstrates the coupling between Gaussian modes and vortex modes between the two waveguides. And the experimental observation can be well understood by expanding the standard coupled mode theory to the vortex-mode case. Here, we present a general model describing the intermode interaction between two waveguides. We follow the representation used in \cite{IEEEJQER.22.988, OER.13.1515} to derive a general model for vortex modes and relevant parameters, which is equivalent to that in \cite{HuangR1994, OkamotoR2006} but is more transparent in physics. This method has been widely used to study the mode coupling between optical resonators \cite{PRAR.75.023814,PRAR.84.013808} and waveguides \cite{OER.21.25619}. 

\renewcommand{\thefigure}{S\arabic{figure}}
\begin{figure*}[htb!]
\centering
\includegraphics[width=0.77\columnwidth]{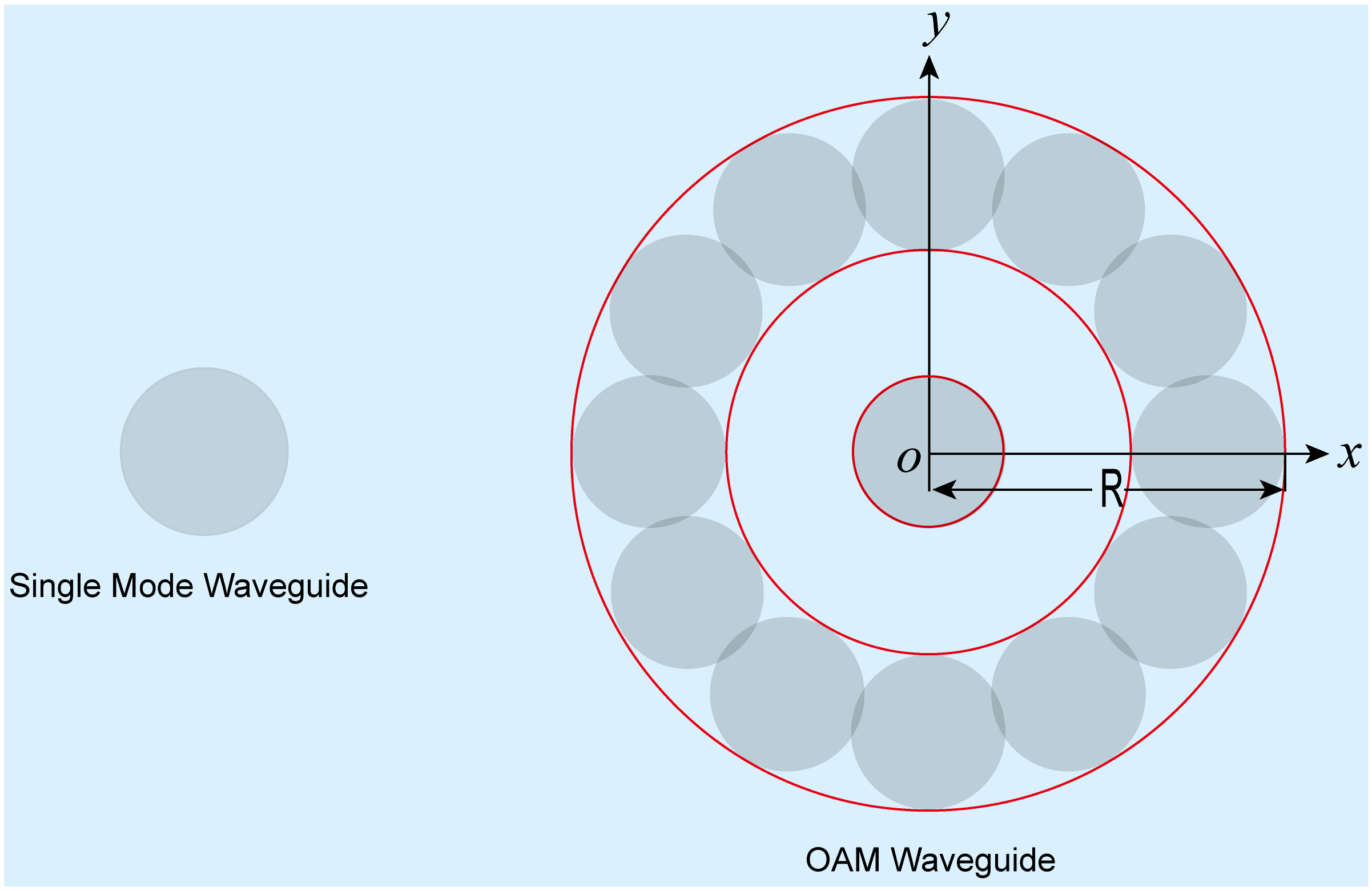}
\caption{\textbf{Cross section of the coupled Gaussian-mode (left) and OAM-mode (right) waveguides}. The light blue square is the wafer material with an electronic constant $\varepsilon_s$. The round gray dots indicates the laser-writing area. The left gray dot is the single-mode waveguide. The right structure depicts the OAM waveguide. The right circles are the profiles of the ideal OAM waveguide to be written.}
\label{Figure S2}
\end{figure*}

\subsection{General Helmholtz equation}
We start our derivation from Maxwell's equations below
\begin{equation}\label{eq:Maxwell}
\begin{split}
\nabla \times {\bf E} & = -\frac{\partial {\bf B}}{\partial t} \;, \\
\nabla \times {\bf H} & = \frac{\partial {\bf D}}{\partial t} \;, \\
\nabla \cdot {\bf D} & = 0 \;, \\
\nabla \cdot {\bf B} & = 0\;.
\end{split}
\end{equation}
Here we consider an optical system that the medium is charge free and current free. The material permittivity and permeability is time independent and $\mu_r =1$. 
 
For a monoharmonic electromangetic (em) field $\{ {\bf E}^\prime = {\bf E} e^{-j\omega t},  {\bf H}^\prime = {\bf H} e^{-j\omega t} \} $ with an angular frequency $\omega$, we have 
 \begin{equation}
 \nabla \times \nabla \times {\bf E} = j\omega \mu_0 \nabla \times {\bf H} = \omega^2 \varepsilon_0 \mu_0 \varepsilon({\bf r}) {\bf E} \;,
 \end{equation}
where $\varepsilon_0$ and $\mu_0$ are the vacuum permittivity and permeability, $\varepsilon({\bf r})$ is the relative permittivity at position ${\bf r}$. The light velocity in the vacuum is $C = 1/\sqrt{\varepsilon_0 \mu_0}$ and $k_0 = \omega/C$ the propagation constant in the free space. We apply the relation $\nabla \times \nabla \times {\bf E} = \nabla (\nabla \cdot {\bf E}) - \nabla^2 {\bf E}$ and obtain
 \begin{equation}\label{eq:Helmholtz}
 \nabla^2 {\bf E} - \nabla (\nabla \cdot {\bf E}) + k_0^2 \varepsilon({\bf r}) {\bf E} = 0 \;.
 \end{equation}
This equation is a general Helmholtz equation for all kinds of waveguides with temporal independent and linear medium. Below, we consider a system consisting of a single-mode waveguide and an OAM-mode waveguide shown in Fig.~\ref{Figure S2}, denoted as the waveguide a and b, respectively. 

\subsection{The electric constant distribution of the waveguides}
The single-mode and OAM-mode waveguides are directly written by laser pulse trains in a wafer made of $\text{SiO}_2$ with an electronic constant $\varepsilon_s$. In ideal case, we want to write a single-mode and an OAM-mode waveguide with Gaussian laser pulses. The effective profiles of the OAM waveguides are illustrated by the red circles. The left one corresponds to the single-mode waveguide. The ideal OAM waveguide is schematically depicted by the three co-central red circles: the small red circle in the middle indicates a high-refractive-index area, while the outside two red circles show a donut-shaped high refractive region. We assume that, if the writing operation is perfect, the profiles of the electric constant distributions for the single-mode and the OAM-mode waveguides are $\varepsilon_s  + \varepsilon_a({\bf r})$ and $\varepsilon_s  + \varepsilon_b ({\bf r})$, respectively, where $\varepsilon_a ({\bf r})$ and $\varepsilon_b({\bf r})$ are the change of electric constant of the designed ideal waveguides with respect to the wafer. 

In practical case, the electric constant distribution of the waveguide is determined by the energy density of the laser pulse and exists fluctuations beyond control in the laser writing process. The deviation of the writing laser energy from an axial symmetrical profile is the main cause resulting in birefringence in the waveguide, leading to the difference of the relative permittivities in two orthogonal directions. We consider an anisotropic imperfection in laser writing as evidenced by Fig.~\ref{Figure S1}. The deviations of the electric constants of two waveguides are $3 \times 3$ tensor, $\Delta {\bf \varepsilon}_a  ({\bf r})$ and $\Delta {\bf \varepsilon}_b ({\bf r})$, respectively. Thus, the total electric constant distribution of the system becomes a tensor and given by
\begin{equation}\label{eq:totalEpsilon}
{\bf \varepsilon} ({\bf r}) = \varepsilon_s + \varepsilon_a  ({\bf r}) + \Delta {\bf \varepsilon}_a  ({\bf r}) + \varepsilon_b  ({\bf r}) + \Delta {\bf \varepsilon}_b  ({\bf r}) \;.
\end{equation}
The single-mode waveguide a and the OAM-mode waveguide b are individually defined by $\varepsilon_s + \varepsilon_a  ({\bf r}) + \Delta {\bf \varepsilon}_a  ({\bf r}) $ and $\varepsilon_s +\varepsilon_b  ({\bf r}) + \Delta {\bf \varepsilon}_b  ({\bf r}) $, respectively. In our case, we have $\varepsilon_s \gg \varepsilon_a  ({\bf r}), \varepsilon_b  ({\bf r}) \gg \Delta {\bf \varepsilon}_a  ({\bf r}), \Delta {\bf \varepsilon}_b  ({\bf r})$.

We consider the waveguide a and b individually and have the Helmholtz equation
 \begin{equation}\label{eq:Helmholtza}
 \nabla^2 {\bf E}_a - \nabla (\nabla \cdot {\bf E}_a) + k_0^2 (\varepsilon_s + \varepsilon_a  ({\bf r}) + \Delta {\bf \varepsilon}_a  ({\bf r})) {\bf E}_a = 0 \;,
 \end{equation}
for the waveguide a, and 
  \begin{equation}\label{eq:Helmholtzb}
 \nabla^2 {\bf E}_b - \nabla (\nabla \cdot {\bf E}_b) + k_0^2 (\varepsilon_s + \varepsilon_b  ({\bf r}) + \Delta {\bf \varepsilon}_b ({\bf r})) {\bf E}_b = 0 \;,
 \end{equation}
for the waveguide b, respectively. If the change of the refractive index $\varepsilon_a ({\bf r}) + \Delta {\bf \varepsilon}_a ({\bf r})$ ($\varepsilon_b ({\bf r}) + \Delta {\bf \varepsilon}_b ({\bf r})$) is anisotropic, then the waveguide eigenmodes polarized along two orthogonal transverse directions, have nondegenerate propagation constants. To make the following analysis simple, we divide the operator $\nabla$ into the transverse component $\nabla_t = \partial_x \hat{x} +  \partial_y \hat{y}$ and the $z$ directional one $\nabla_z = \partial_z \hat{z}$ that $\nabla^2  = \nabla_t^2 + \nabla_z^2$ with $\nabla_t^2 = \partial_x^2 + \partial_y^2$ and $\nabla_z^2 = \partial_z^2$.

\subsection{Coupling between eigenmodes in two waveguides}
An optical waveguide supports some eigenmodes, whose transverse distributions of fields are determined by the transverse profile of the refractive index. Once the eigenmodes are found, the electric field propagating in the waveguide can be considered as a superposition of eigenmodes with different weights. Next, beginning with the general Helmholtz equation Eq.~\ref{eq:Helmholtz}, we derive the mode-coupling equations. 

We write the $m$th ($n$th) eigenmode ${\bf E}_m$ (${\bf E}_n$) of the waveguide a (b) without the perturbation $\Delta \varepsilon_a$ ($\Delta\varepsilon_b$) in the form of ${\bf E}_m = \tilde{\bf E}_m(x,y) e^{-j\beta_m z}$ [${\bf E}_n = \tilde{\bf E}_n(x,y) e^{-j\beta_n z}$] to separate the field into the transverse profile $\tilde{\bf E}_m(x,y)$ and the $z$ dependence, where $\beta_m$ ($\beta_n$) is the propagation constant. Then, we have the Helmholtz equation for the eigenmode of the waveguide a and b individually
\begin{equation}\label{eq:HelmholtzEngp}
 \nabla^2 {\bf E}_m - \nabla (\nabla \cdot {\bf E}_m) + k_0^2 (\varepsilon_s + \varepsilon_a  ({\bf r}) ) {\bf E}_m = 0 \;,
 \end{equation}
for the waveguide a, and 
  \begin{equation}\label{eq:HelmholtzEngb}
 \nabla^2 {\bf E}_n - \nabla (\nabla \cdot {\bf E}_n) + k_0^2 (\varepsilon_s + \varepsilon_b  ({\bf r}) ) {\bf E}_n = 0 \;,
 \end{equation}
for the waveguide b. These eigenmodes in the two waveguides a and b are orthogonal with each other that 
 \begin{subequations}
 \begin{align}
 \int \tilde{\mathbf E}_m \cdot \tilde{\mathbf E}^*_{m^\prime} dx dy &= N^a_m \delta (m - m^\prime)\;,
 \\
 \int \tilde{\mathbf E}_n \cdot \tilde{\mathbf E}^*_{n^\prime} dx dy &= N^b_n \delta (n - n^\prime) \;.
 \end{align}
 \end{subequations}
 %
 Using the relation 
 \begin{equation}
 \nabla^2 \tilde{\bf E}_p e^{-j\beta_p z} = e^{-j\beta_p z} \nabla^2 \tilde{\bf E}_p - \beta_p^2 e^{-j\beta_p z} \tilde{\bf E}_p \;,
 \end{equation}
  we can obtain 
 \begin{subequations}\label{eq:1}
 \begin{align}
 & e^{-j\beta_m z}  \nabla^2 \tilde{\bf E}_m - \beta_m^2 e^{-j\beta_m z}   \tilde{\bf E}_m - \nabla \left( \nabla \cdot \tilde{\bf E}_m e^{-j\beta_m z}  \right) + k_0^2 (\varepsilon_s +\varepsilon_a({\bf r})) \tilde{\bf E}_m e^{-j\beta_m z}  =0 \;, \\
 & e^{-j\beta_n z}  \nabla^2 \tilde{\bf E}_n - \beta_n^2 e^{-j\beta_n z}   \tilde{\bf E}_n - \nabla \left( \nabla \cdot \tilde{\bf E}_n e^{-j\beta_n z}  \right) + k_0^2 (\varepsilon_s +\varepsilon_b({\bf r})) \tilde{\bf E}_n e^{-j\beta_n z}  =0 \;. 
 \end{align}
 \end{subequations}
We consider a field ${\bf E}$ when the two waveguides a and b coexist, as shown in Fig.~\ref{Figure S2} that the profile of the refractive index $\varepsilon({\bf r})$ is given by Eq.~\ref{eq:totalEpsilon}. Then, the field is the superposition of the eigenmodes and can be written as
\begin{equation}\label{eq:totalE}
{\bf E} = \sum_m A_m \tilde{\bf E}_m e^{-j\beta_m z} + \sum_n B_n \tilde{\bf E}_n e^{-j\beta_n z} \;,
\end{equation}
where $A_m$ and $B_n$ are the amplitudes of the corresponding eigenmodes $\tilde{\bf E}_m$ and $\tilde{\bf E}_n$ in the waveguide a and b, respectively. They are only $z$ dependent, i.e. $\nabla A_m = \frac{\partial A_m}{\partial z} \cdot \hat{z}$. The propagation of the field ${\bf E}$ in Eq.~\ref{eq:totalE} is governed by Eq.~\ref{eq:Helmholtz}. Substituting Eq.~\ref{eq:totalE}  into Eq.~\ref{eq:Helmholtz}, we obtain 
\begin{equation} \label{eq:HelmholtzEng}
\begin{split}
& \sum_m \nabla^2 A_m \tilde{\bf E}_m e^{-j\beta_m z} - \sum_m \nabla \left( \nabla \cdot A_m \tilde{\bf E}_m e^{-j\beta_m z} \right) + \sum_m k_0^2 \varepsilon ({\bf r}) A_m \tilde{\bf E}_m e^{-j\beta_m z} \\
& + \sum_n \nabla^2 B_n \tilde{\bf E}_n e^{-j\beta_n z} - \sum_n \nabla \left( \nabla \cdot B_n \tilde{\bf E}_n e^{-j\beta_n z} \right) + \sum_n k_0^2 \varepsilon({\bf r}) B_n \tilde{\bf E}_n e^{-j\beta_n z} =0 \;.
\end{split}
\end{equation}
 
 and
 \begin{subequations} \label{eq:2}
 \begin{align}
 & \nabla^2 A_m \tilde{\mathbf E}_m e^{-j\beta_m z} = \tilde{\mathbf E}_m e^{-j\beta_m z} \frac{\partial^2 A_m}{\partial z^2} -2 j\beta_m e^{-j\beta_m z}  \frac{\partial A_m}{\partial z}  \tilde{\mathbf E}_m +A_m e^{-j\beta_m z}  \nabla^2 \tilde{\mathbf E}_m - A_m \beta_m^2 e^{-j\beta_m z} \tilde{\mathbf E}_m \;,\\
  & \nabla^2 B_n \tilde{\mathbf E}_n e^{-j\beta_n z} = \tilde{\mathbf E}_n e^{-j\beta_n z} \frac{\partial^2 B_n}{\partial z^2} -2 j\beta_n e^{-j\beta_n z}  \frac{\partial B_n}{\partial z}  \tilde{\mathbf E}_n +B_n e^{-j\beta_n z}  \nabla^2 \tilde{\mathbf E}_n - B_n \beta_n^2 e^{-j\beta_n z} \tilde{\mathbf E}_n \;.
 \end{align}
 \end{subequations}
We also have 
\begin{subequations} \label{eq:3}
 \begin{align}
 \nabla \cdot A_m \tilde{\mathbf E}_m e^{-j\beta_m z} & = A_m \nabla\cdot \tilde{\mathbf E}_m e^{-j\beta_m z} + e^{-j\beta_m z}  \frac{\partial A_m}{\partial z} \tilde{\mathbf E}_m \cdot \hat{z} \;,\\
 \nabla \cdot B_n \tilde{\mathbf E}_n e^{-j\beta_n z} & = B_n \nabla\cdot \tilde{\mathbf E}_n e^{-j\beta_n z} + e^{-j\beta_n z}  \frac{\partial B_n}{\partial z} \tilde{\mathbf E}_n \cdot \hat{z} \;,\\
 \begin{split}
  \nabla \left(  \nabla \cdot A_m \tilde{\mathbf E}_m e^{-j\beta_m z}  \right) & = e^{-j\beta_m z} \frac{\partial A_m}{\partial z} \left( \nabla \cdot \tilde{\mathbf E}_m \right) \hat{z} - j\beta_m e^{-j\beta_m z} \frac{\partial A_m}{\partial z} \tilde{E}_m^z \hat{z} + A_m \nabla \left(\nabla \cdot \tilde{\mathbf E}_m e^{-j\beta_m z} \right) \\
  &   
  + e^{-j\beta_m z} \frac{\partial A_m}{\partial z}\nabla \tilde{E}_m^z  -j \beta_m e^{-j\beta_m z} \frac{\partial A_m}{\partial z}  \tilde{E}_m^z \hat{z} + e^{-j\beta_m z} \frac{\partial^2 A_m}{\partial z^2}  \tilde{E}_m^z \hat{z} \;,
  \end{split}\\
   \begin{split}
  \nabla \left(  \nabla \cdot B_n \tilde{\mathbf E}_n e^{-j\beta_n z}  \right) & = e^{-j\beta_n z} \frac{\partial B_n}{\partial z} \left( \nabla \cdot \tilde{\mathbf E}_n \right) \hat{z} - j\beta_n e^{-j\beta_n z} \frac{\partial B_n}{\partial z} \tilde{E}_n^z \hat{z} + B_n \nabla \left(\nabla \cdot \tilde{\mathbf E}_n e^{-j\beta_n z} \right) \\
  &   
  + e^{-j\beta_n z} \frac{\partial B_n}{\partial z}\nabla \tilde{E}_n^z  -j \beta_n e^{-j\beta_n z} \frac{\partial B_n}{\partial z}  \tilde{E}_n^z \hat{z} + e^{-j\beta_n z} \frac{\partial^2 B_n}{\partial z^2}  \tilde{E}_n^z \hat{z} \;,
  \end{split}
\end{align}
 \end{subequations}
 with $\tilde{E}_m^z = \tilde{\mathbf E}_m \cdot \hat{z}$ and $\tilde{E}_n^z = \tilde{\mathbf E}_n \cdot \hat{z}$.
 
 
Substituting Eqs.~\ref{eq:2} and ~\ref{eq:3} into Eq.~\ref{eq:HelmholtzEng}, we get
\begin{equation} \label{eq:ExpdE}
\begin{split}
& \sum_m (\tilde{\mathbf E}_m - \tilde{E}_m^z \hat{z}) e^{-j\beta_m z} \frac{\partial^2 A_m}{\partial z^2} + \sum_m  A_m k_0^2 \left(\Delta\varepsilon_a + \varepsilon_b + \Delta\varepsilon_b \right) \tilde{\mathbf E}_m e^{-j\beta_m z}\\
&+ \sum_m \left[-2j\beta_m \frac{\partial A_m}{\partial z} (\tilde{\mathbf E}_m - \tilde{E}_m^z \hat{z})  - \left( \nabla \tilde{E}_m^z + \nabla \cdot \tilde{\mathbf E}_m \hat{z}\right)   \frac{\partial A_m}{\partial z}\right] e^{-j\beta_m z}\\
& + \sum_m A_m \left[e^{-j\beta_m z} \nabla^2 \tilde{\mathbf E}_m - \beta_m^2 \tilde{\mathbf E}_m e^{-j\beta_m z}  - \nabla\left(\nabla \cdot  \tilde{\mathbf E}_m e^{-j\beta_m z}  \right) + k_0^2 (\varepsilon_s + \varepsilon_a) \tilde{\mathbf E}_m e^{-j\beta_m z}  \right] \\
& \sum_n (\tilde{\mathbf E}_n - \tilde{E}_n^z \hat{z}) e^{-j\beta_n z} \frac{\partial^2 B_n}{\partial z^2} + \sum_n  B_n k_0^2 \left(\Delta\varepsilon_b + \varepsilon_a + \Delta\varepsilon_a \right) \tilde{\mathbf E}_n e^{-j\beta_n z}\\
&+ \sum_n \left[-2j\beta_n \frac{\partial B_n}{\partial z} (\tilde{\mathbf E}_n - \tilde{E}_n^z \hat{z})  -\left( \nabla \tilde{E}_n^z + \nabla \cdot \tilde{\mathbf E}_n \hat{z}\right)   \frac{\partial B_n}{\partial z} \right]e^{-j\beta_n z} \\
& + \sum_n B_n \left[e^{-j\beta_n z} \nabla^2 \tilde{\mathbf E}_n - \beta_n^2 \tilde{\mathbf E}_n e^{-j\beta_n z}  - \nabla\left(\nabla \cdot  \tilde{\mathbf E}_n e^{-j\beta_n z}  \right) + k_0^2 (\varepsilon_s + \varepsilon_b) \tilde{\mathbf E}_n e^{-j\beta_n z}  \right] =0 \;.
\end{split}
\end{equation}
Normally, we have $\left| \tilde{\mathbf E}_p \cdot \hat{z} \right| \ll \left| \tilde{\mathbf E}_p \cdot (\hat{x} +  \hat{y})\right| \approx \left| \tilde{\mathbf E}_p\right|$ and $\beta_p \gg \left| \left( \nabla \tilde{E}_p^z + \nabla \cdot \tilde{\mathbf E}_p \hat{z}\right)   \frac{\partial A_m}{\partial z} \right|$ with $p=m,n$. Thus, we  get the approximation

\begin{equation} \label{eq:ExpdE}
\begin{split}
& \sum_m \tilde{\mathbf E}_m e^{-j\beta_m z} \frac{\partial^2 A_m}{\partial z^2} + \sum_m \left[-2j\beta_m \frac{\partial A_m}{\partial z} \tilde{\mathbf E}_m e^{-j\beta_m z} + A_m k_0^2 \left(\Delta\varepsilon_a + \varepsilon_b + \Delta\varepsilon_b \right) \tilde{\mathbf E}_m e^{-j\beta_m z} \right] \\
& + \sum_m A_m \left[e^{-j\beta_m z} \nabla^2 \tilde{\mathbf E}_m - \beta_m^2 \tilde{\mathbf E}_m e^{-j\beta_m z}  - \nabla\left(\nabla \cdot  \tilde{\mathbf E}_m e^{-j\beta_m z}  \right) + k_0^2 (\varepsilon_s + \varepsilon_a) \tilde{\mathbf E}_m e^{-j\beta_m z}  \right] \\
+ & \sum_n \tilde{\mathbf E}_n e^{-j\beta_n z} \frac{\partial^2 B_n}{\partial z^2} +\sum_n \left[-2j\beta_n \frac{\partial B_n}{\partial z} \tilde{\mathbf E}_n e^{-j\beta_n z} + B_n k_0^2 \left(\Delta\varepsilon_b + \varepsilon_a + \Delta\varepsilon_a \right) \tilde{\mathbf E}_n e^{-j\beta_n z} \right] \\
& + \sum_n B_n \left[e^{-j\beta_n z} \nabla^2 \tilde{\mathbf E}_n - \beta_n^2 \tilde{\mathbf E}_n e^{-j\beta_n z}  - \nabla\left(\nabla \cdot  \tilde{\mathbf E}_n e^{-j\beta_n z}  \right) + k_0^2 (\varepsilon_s + \varepsilon_b) \tilde{\mathbf E}_n e^{-j\beta_n z}  \right]=0 \;.
\end{split}
\end{equation}
Because of Eq.~\ref{eq:1}, the third and the last terms are zero. Applying the approximations $\left| \frac{\partial^2 A_m}{\partial z^2}\right| \ll \left| 2\beta_m \frac{\partial A_m}{\partial z}\right|$ and $\left| \frac{\partial^2 B_n}{\partial z^2}\right| \ll \left| 2\beta_n \frac{\partial B_n}{\partial z}\right|$, and neglecting the terms relevant to $\frac{\partial^2 A_m}{\partial z^2}$ and $\frac{\partial^2 B_n}{\partial z^2}$, we obtain the final equation describing the coupling of the eigenmodes as 
\begin{equation} \label{eq:CoupledEmEn}
\begin{split}
& \sum_m \left[-2j\beta_m \frac{\partial A_m}{\partial z} \tilde{\mathbf E}_m  + A_m k_0^2 \left(\Delta\varepsilon_a + \varepsilon_b + \Delta\varepsilon_b \right) \tilde{\mathbf E}_m  \right] e^{-j\beta_m z} \\
& + \sum_n \left[-2j\beta_n \frac{\partial B_n}{\partial z} \tilde{\mathbf E}_n + B_n k_0^2 \left(\Delta\varepsilon_b + \varepsilon_a + \Delta\varepsilon_a \right) \tilde{\mathbf E}_n  \right] e^{-j\beta_n z} =0 \;.
\end{split}
\end{equation}
Note that $\Delta\varepsilon_a$ and $\Delta\varepsilon_b$ are the tensors of dielectric constant fluctuations.

Multipling Eq.~\ref{eq:CoupledEmEn} with $\tilde{\mathbf{E}}_{m^\prime}^*$ and then integrating over the cross section, i.e. $\int \tilde{\mathbf E}_{m^\prime}^* \cdot \text{Eq.~18}~ dxdy$, we get
\begin{equation}
  \begin{split}
      2j\beta_{m^\prime} N_{m^\prime} e^{-j\beta_{m^\prime}z} \frac{\partial A_{m^\prime}}{\partial z} = & \sum_m k_0^2 \int \tilde{\mathbf{E}}_{m^\prime}^* \cdot (\varepsilon_b + \Delta\varepsilon_a + \Delta\varepsilon_b) \cdot \tilde{\mathbf{E}}_m dx dy ~ e^{-j\beta_{m}z} A_m \; \\
      & -2j \sum_n \beta_n \int \tilde{\mathbf{E}}_{m^\prime}^* \cdot  \tilde{\mathbf E}_ndx dy ~ e^{-j\beta_{n}z}  \frac{\partial B_n}{\partial z} \;\\
      & + \sum_n  k_0^2 \int \tilde{\mathbf{E}}_{m^\prime}^* \cdot (\varepsilon_a + \Delta\varepsilon_a + \Delta\varepsilon_b) \cdot \tilde{\mathbf{E}}_n dx dy ~ e^{-j\beta_{n}z} B_n \;.
  \end{split}
\end{equation}
This equation can be rewritten as
\begin{equation} \label{eq:Am}
      \frac{\partial A_{m^\prime}}{\partial z} =  -j \delta_{m^\prime} A_{m^\prime} - j \sum_{m \neq m^\prime} h^{(a)}_{m,m^\prime} e^{-j\Delta \beta_{m,m^\prime}z}  A_m 
        - \sum_n  C_{n,m^\prime} e^{-j\Delta \beta_{n,m^\prime}z}   \frac{\partial B_n}{\partial z} -j \sum_n \kappa_{n,m^\prime} e^{-j\Delta \beta_{n,m^\prime}z}  B_n \;,
\end{equation}
with 
\begin{subequations}
\begin{align}
\Delta \beta_{m,m^\prime} = & \beta_m -\beta_{m^\prime} \;,\\
\Delta \beta_{n,m^\prime} = & \beta_n -\beta_{m^\prime} \;,\\
 \delta_{m^\prime} = & \frac{k_0^2}{2 \beta_{m^\prime} N_{m^\prime}} \int \tilde{\mathbf{E}}_{m^\prime}^* \cdot (\varepsilon_b + \Delta\varepsilon_b + \Delta\varepsilon_a) \cdot \tilde{\mathbf{E}}_{m^\prime} dx dy \;, \\
 h^{(a)}_{m,m^\prime} = &  \frac{k_0^2}{2 \beta_{m^\prime} N_{m^\prime}} \int \tilde{\mathbf{E}}_{m^\prime}^* \cdot (\varepsilon_b + \Delta\varepsilon_b + \Delta\varepsilon_a) \cdot \tilde{\mathbf{E}}_m dx dy  \;, \\
 C_{n,m^\prime} = & \frac{\beta_n}{\beta_{m^\prime} N_{m^\prime}} \int \tilde{\mathbf{E}}_{m^\prime}^* \cdot  \tilde{\mathbf E}_ndx dy \;,\\
 \kappa_{n,m^\prime} = &  \frac{k_0^2}{2 \beta_{m^\prime} N_{m^\prime}} \int \tilde{\mathbf{E}}_{m^\prime}^* \cdot (\varepsilon_a + \Delta\varepsilon_a + \Delta\varepsilon_b) \cdot \tilde{\mathbf{E}}_n dx dy  \;, 
\end{align}
\end{subequations}
where $\Delta \beta_{m,m^\prime}$ ($\Delta \beta_{n,m^\prime}$) is the propagation constant mismatching between the $m$th ($n$th) eigenmode in the waveguide a (b) and the $m^\prime$th eigenmode in the waveguide a. The fluctuation $\Delta\varepsilon_a$ and the existance of the waveguide b cause the propagation constant shift $\delta_{m^\prime}$  to the $m^\prime$ mode, the coupling $ h^{(a)}_{m,m^\prime}$ between the $m$th and $m^\prime$th eigenmodes in the waveguide a, and the directional coupling $\kappa_{n,m^\prime}$ between the waveguide a and b. 

Similarly, we multiple Eq.~\ref{eq:CoupledEmEn} with $\tilde{\mathbf{E}}_{n^\prime}^*$ and integrate over the cross section, then we can get
\begin{equation} \label{eq:Bn}
      \frac{\partial B_{n^\prime}}{\partial z} =  -j \delta_{n^\prime} B_{n^\prime} - j \sum_{n \neq n^\prime} h^{(b)}_{n,n^\prime} e^{-j\Delta \beta_{n,n^\prime}z}  B_n 
        - \sum_m C_{m,n^\prime} e^{-j\Delta \beta_{m,n^\prime}z}   \frac{\partial A_m}{\partial z} -j \sum_m \kappa_{m,n^\prime} e^{-j\Delta \beta_{m,n^\prime}z}  A_m \;,
\end{equation}
with 
\begin{subequations}
\begin{align}
\Delta \beta_{n,n^\prime} = & \beta_n -\beta_{n^\prime} \;,\\
\Delta \beta_{m,n^\prime} = & \beta_m -\beta_{n^\prime} \;,\\
 \delta_{n^\prime} = & \frac{k_0^2}{2 \beta_{n^\prime} N_{n^\prime}} \int \tilde{\mathbf{E}}_{n^\prime}^* \cdot (\varepsilon_a + \Delta\varepsilon_a + \Delta\varepsilon_b) \cdot \tilde{\mathbf{E}}_{n^\prime} dx dy \;, \\
 h^{(b)}_{n,n^\prime} = &  \frac{k_0^2}{2 \beta_{n^\prime} N_{n^\prime}} \int \tilde{\mathbf{E}}_{n^\prime}^* \cdot (\varepsilon_a + \Delta\varepsilon_a + \Delta\varepsilon_b) \cdot \tilde{\mathbf{E}}_n dx dy  \;, \\
 C_{m,n^\prime} = & \frac{\beta_m}{\beta_{n^\prime} N_{n^\prime}} \int \tilde{\mathbf{E}}_{n^\prime}^* \cdot  \tilde{\mathbf E}_m dx dy \;,\\
 \kappa_{m,n^\prime} = &  \frac{k_0^2}{2 \beta_{n^\prime} N_{n^\prime}} \int \tilde{\mathbf{E}}_{n^\prime}^* \cdot (\varepsilon_b + \Delta\varepsilon_b + \Delta\varepsilon_a) \cdot \tilde{\mathbf{E}}_m dx dy  \;, 
\end{align}
\end{subequations}
where $\Delta \beta_{n,n^\prime}$ ($\Delta \beta_{m,n^\prime}$) is the propagation constant mismatching between the $n$th ($m$th) eigenmode in the waveguide b (a) and the $n^\prime$th eigenmode in the waveguide b. The fluctuation $\Delta\varepsilon_b$ and the existance of the waveguide a cause the propagation constant shift $\delta_{n^\prime}$  to the $n^\prime$ mode, the coupling $ h^{(b)}_{n,n^\prime}$ between the $n$th and $n^\prime$th eigenmodes in the waveguide b, and the directional coupling $\kappa_{m,n^\prime}$ between the waveguide a and b. The coupling $C_{n,m^\prime}$ ($C_{m,n^\prime}$) between the eigenmodes in the two waveguides is the so-called Butt coupling, simply because of the small overlap in space between two well seperate modes. Note that the propagation constants of all eigenmodes are close to each other that $\beta_m \sim \beta_{m^\prime} \sim k_0$ and $\beta_n \sim \beta_{n^\prime} \sim k_0$, about $8\times 10^6$ rad/m. In our experiments, $\varepsilon_a$ and $\varepsilon_b$ is the order of $10^{-3}\varepsilon_s$. The integral in the Butt couplings is about $3$ orders larger than those in the shifts and the intermode couplings. Thus, the Butt couplings $C_{n,m^\prime}$ and  $C_{m,n^\prime}$ can be neglected because they are three orders smaller than the shifts $\delta_{m^\prime}$ and $\delta_{n^\prime}$, the intermode couplings $ h^{(a)}_{m,m^\prime}$, $ h^{(b)}_{n,n^\prime}$, $ \kappa_{n,m^\prime}$ and $ \kappa_{m,n^\prime}$. When the Butt couplings are negligible, we have $\kappa_{n,m} = \kappa_{m,n}^*$ \cite{OkamotoR2006, IEEEJQER.22.988, JLTR.5.16, IEEEJQER.23.1689}. If we neglect the absorption of material, then $\kappa_{n,m} = \kappa_{m,n}$.
According to Eqs.~28 and 30, the propagation constant shift $\delta_{m^\prime}$ ($\delta_{n^\prime}$), and the intrawaveguide coupling $h^{(a)}_{m,m^\prime}$ ($h^{(b)}_{n,n^\prime}$) are resulted from the overlapping of leakage of two exponentially decaying evanescent fields $\tilde{\bf E}_{m^\prime}$ and $\tilde{\bf E}_{m}$ ($\tilde{\bf E}_{n^\prime}$ and $\tilde{\bf E}_{n}$) to the neighbouring OAM (single-mode) waveguide. In contrast, the interwaveguide coupling $\kappa_{n,m^\prime}$ ($\kappa_{m,n^\prime}$) is resulted from the overlapping of the field $\tilde{\bf E}_{m^\prime}$ ($\tilde{\bf E}_{n^\prime}$) in the single-mode (OAM) waveguide and the leakage of the eigenmode field $\tilde{\bf E}_{n}$ ($\tilde{\bf E}_{m}$) of the neighbouring OAM (single-mode) waveguide. Because the evanescent component of an eigenmode field in the neighbouring waveguide is much smaller than its component in the supporting waveguide, the interwaveguide coupling is typically much larger than the shifts and the intrawaveguide coupling. The propagation constant shifts and the intermode couplings in Eq.~\ref{eq:Am} and Eq.~\ref{eq:Bn} include two contributions from: the dielectric changes due to the existence of the neighbouring ideal waveguide; the dielectric fluctuations in the two waveguides.

As mentioned above, the deviation of the writing laser energy from an axial symmetrical profile is the main cause resulting in birefringence in the waveguide, leading to the difference of the relative permittivities in two orthogonal directions. Therefore, the 
relative permittivities can be written as 
\begin{subequations} \label{eq:relativepermittivities}
\begin{align}
\Delta\varepsilon_{a}=& \begin{pmatrix} \Delta\varepsilon_{x^\prime}& 0& 0\\ 0 & \Delta\varepsilon_{y^\prime} & 0\\ 0& 0& \Delta\varepsilon_{z}\end{pmatrix} \;,\\
\Delta\varepsilon_{b}=& \begin{pmatrix} \Delta\varepsilon_{x}& 0& 0\\ 0 & \Delta\varepsilon_{y} & 0\\ 0& 0& \Delta\varepsilon_{z} \end{pmatrix} \;,
\end{align}
\end{subequations}
where $\Delta\varepsilon_{x^\prime}\ne \Delta\varepsilon_{y^\prime}$, $\Delta\varepsilon_{x}\ne \Delta\varepsilon_{y}$. The propagation constant shift $\delta_{m^\prime}$ ($\delta_{n^\prime}$), the coupling $ h^{(a)}_{m,m^\prime}$ ($h^{(b)}_{n,n^\prime}$) between the $m$th ($n$th) and $m^\prime$th ($n^\prime$th) eigenmodes in the waveguide a (b), and the directional coupling $\kappa_{n,m^\prime}$ ($\kappa_{m,n^\prime}$) between the waveguides a and b are all related to birefringence of the waveguide.

The optical axes of the two waveguides are not necessary parallel. In deed, they normally have small angles. The rotation transformation matrix $R(\theta)$ of the optical axes between two waveguides satisfies the following relationship:
\begin{equation} \label{eq1:matrixtransformation}
\begin{pmatrix} {\mathbf e}^a_{x^\prime} \\ {\mathbf e}^a_{y^\prime}\\ {\mathbf e}^a_z \end{pmatrix}=R(\theta)\begin{pmatrix} {\mathbf e}^b_x \\ {\mathbf e}^b_y \\ {\mathbf e}^b_z\end{pmatrix}= \begin{pmatrix} \cos\theta& -\sin\theta & 0\\ \sin\theta & \cos\theta & 0\\ 0& 0& 1\end{pmatrix} \begin{pmatrix} {\mathbf e}^b_x \\ {\mathbf e}^b_y \\ {\mathbf e}^b_z\end{pmatrix} \;,
\end{equation}
where $\theta$ is the rotation angle between the optical axes of two waveguides. When $\theta=0$, the rotation transformation matrix $R(\theta)$ is an identity matrix. Neglecting the Butt couplings $C_{n,m^\prime}$ and $C_{m,n^\prime}$ and considering a nozero $\theta$, we can rewrite the coefficients of Eq.~\ref{eq:Am} and Eq.~\ref{eq:Bn} as 
\begin{subequations}
\begin{align}
 \delta_{m^\prime} = & \frac{k_0^2}{2 \beta_{m^\prime} N_{m^\prime}} \int \tilde{\mathbf{E}}_{m^\prime}^* \cdot (\varepsilon_b + R(\theta)\Delta\varepsilon_b + \Delta\varepsilon_a) \cdot \tilde{\mathbf{E}}_{m^\prime} dx dy \;, \\
 h^{(a)}_{m,m^\prime} = &  \frac{k_0^2}{2 \beta_{m^\prime} N_{m^\prime}} \int \tilde{\mathbf{E}}_{m^\prime}^* \cdot (\varepsilon_b + R(\theta)\Delta\varepsilon_b + \Delta\varepsilon_a) \cdot \tilde{\mathbf{E}}_m dx dy  \;, \\
 \kappa_{n,m^\prime} = &  \frac{k_0^2}{2 \beta_{m^\prime} N_{m^\prime}} \int \tilde{\mathbf{E}}_{m^\prime}^* \cdot (\varepsilon_a + \Delta\varepsilon_a + R(\theta)\Delta\varepsilon_b) \cdot R(\theta)\tilde{\mathbf{E}}_n dx dy  \;, \\
  \delta_{n^\prime} = & \frac{k_0^2}{2 \beta_{n^\prime} N_{n^\prime}} \int \tilde{\mathbf{E}}_{n^\prime}^* \cdot (\varepsilon_a + R^{-1}(\theta)\Delta\varepsilon_a+\Delta\varepsilon_b) \cdot \tilde{\mathbf{E}}_{n^\prime}dx dy \;, \\
 h^{(b)}_{n,n^\prime} =&  \frac{k_0^2}{2 \beta_{n^\prime} N_{n^\prime}} \int \tilde{\mathbf{E}}_{n^\prime}^* \cdot (\varepsilon_a + R^{-1}(\theta)\Delta\varepsilon_a+\Delta\varepsilon_b) \cdot \tilde{\mathbf{E}}_{n}dx dy \;, \\
 \kappa_{m,n^\prime} = &  \frac{k_0^2}{2 \beta_{n^\prime} N_{n^\prime}} \int \tilde{\mathbf{E}}_{n^\prime}^* \cdot (\varepsilon_b + \Delta\varepsilon_b + R^{-1}(\theta)\Delta\varepsilon_a) \cdot R^{-1}(\theta)\tilde{\mathbf{E}}_m dx dy  \;. 
\end{align}
\end{subequations}
A small angle can cause the cross coupling between modes with orthogonal polarizations. 

Further, we assume that $X=e^{j(\beta_{X}-\bar{\beta})z}\tilde{X}$, where $\bar{\beta}$ means average propagation constant. Then the above mode amplitudes $A_{m^\prime}=e^{j(\beta_{m^\prime}-\bar{\beta})z}\tilde{A}_{m^\prime}$, $B_{n^\prime}=e^{j(\beta_{n^\prime}-\bar{\beta})z}\tilde{B}_{n^\prime}$. After this transformation, the expansion of the field in eigenmodes takes the form
\begin{equation}
\tilde{\bf E} = \sum_m \tilde{A}_m \tilde{\bf E}_m + \sum_n \tilde{B}_n \tilde{\bf E}_n \;,
\end{equation}
and ${\bf E} = \tilde{\bf E} e^{-j\bar\beta z} $. We define $\Delta{\beta_{m^{\prime}}}=\beta_{m^{\prime}}-\bar{\beta}+ \delta_{m^{\prime}}$ and $\Delta{\beta_{n^{\prime}}}=\beta_{n^{\prime}}-\bar{\beta}+ \delta_{n^{\prime}}$, then the above general coupled mode equations Eq.~\ref{eq:Am} and Eq.~\ref{eq:Bn} can be rewritten as 
\begin{subequations} \label{eq:GCME}
  \begin{align}
     \frac{\partial \tilde{A}_{m^\prime}}{\partial z} = &  -j\Delta{\beta_{m^{\prime}}}\tilde{A}_{m^\prime}- j \sum_{m \neq m^\prime} h^{(a)}_{m,m^\prime}  \tilde{A}_m  -j \sum_n \kappa_{n,m^\prime} \tilde{B}_n \;,\\
     \frac{\partial \tilde{B}_{n^\prime}}{\partial z} = & -j\Delta{\beta_{n^{\prime}}}\tilde{B}_{n^\prime}- j \sum_{n \neq n^\prime} h^{(b)}_{n,n^\prime} \tilde{B}_n -j \sum_m \kappa_{m,n^\prime} \tilde{A}_m\;.
  \end{align}
\end{subequations}
Idealy, if we can write the waveguides perfectly without fluctuation, i.e. $\Delta\varepsilon_a =0$ and $\Delta\varepsilon_b =0$, that all modes have the same propagation constants, then $\Delta{\beta_{m^{\prime}}}=\delta_{m^{\prime}}$ and $\Delta{\beta_{n^{\prime}}}=\delta_{n^{\prime}}$.

\subsection{Coupling modes equations including six modes}
In our experiments, we consider six modes in each case: the two orthogonal eigenmodes $E^a_0 {\mathbf e}^a_{x^\prime}$ and $E^a_0 {\mathbf e}^a_{y^\prime}$ in the single-mode waveguide a and the four vortex eigenmodes $E^b_0 e^{i \ell \phi} {\mathbf e}^b_x$ and $E^b_0 e^{i \ell \phi} {\mathbf e}^b_y$ with $\ell = \pm 1$ or $\ell = \pm 2$ in the OAM waveguide b. The anisotropic fluctuations $\Delta\varepsilon_a$ and $\Delta\varepsilon_b$ induce the birefringence in the waveguides and define the two orthogonal optical axes. The birefringence originates from the substrate material and the irradiation conditions: the asymmetric profile of the waveguide cross-section, mechanical stress in the modified region or laser-induced intrinsic birefringence aligned according to the writing beam polarization \cite{CorrielliR2014}. In our experiments, the focused region of writing laser is elliptical even after beam shaping, inevitably inducing birefringence. 

Taking into account the waveguide structure, $G_{x^\prime}$ and $G_{y^\prime}$ represent the Gaussian eigenmodes along the birefringent axis $x^\prime$ and $y^\prime$ in the single-mode waveguide. Similarily, $x_{\ell}$, $x_{-\ell}$, $y_{\ell}$ and $y_{-\ell}$ represent the eignmodes in the birefringent OAM-waveguide, that is, twisted light carrying $\ell \hbar$ or $-\ell \hbar$ orbital angular momentum along the axis $x$ and $y$. Then, the general coupled mode equations Eqs.~\ref{eq:GCME} can be further reduced into specific coupled mode equations with six eigenmodes including $G_{x^\prime}$, $G_{y^\prime}$, $x_{\ell}$, $x_{-\ell}$, $y_{\ell}$, $y_{-\ell}$.
\begin{subequations} \label{eq:Sixmodes}
  \begin{align}
     \frac{\partial \tilde{A}_{G_{x^\prime}}}{\partial z} =&- j \big[\Delta\beta_{G_{x^\prime}} \tilde{A}_{G_{x^\prime}}+h^{(a)}_{G_{y^\prime},G_{x^\prime}} \tilde{A}_{G_{y^\prime}}+\kappa_{x_{\ell},G_{x^\prime}} \tilde{B}_{x_{\ell}}+ \kappa_{x_{-\ell},G_{x^\prime}} \tilde{B}_{x_{-\ell}}+ \kappa_{y_{\ell},G_{x^\prime}} \tilde{B}_{y_{\ell}}+\kappa_{y_{-\ell},G_{x^\prime}} \tilde{B}_{y_{-\ell}}\big] \;, \\
\frac{\partial \tilde{A}_{G_{y^\prime}}}{\partial z} =&- j\big[h^{(a)}_{G_{x^\prime},G_{y^\prime}} \tilde{A}_{G_{x^\prime}}+\Delta\beta_{G_{y^\prime}} \tilde{A}_{G_{y^\prime}}+\kappa_{x_{\ell},G_{y^\prime}} \tilde{B}_{x_{\ell}}+ \kappa_{x_{-\ell},G_{y^\prime}} \tilde{B}_{x_{-\ell}}+\kappa_{y_{\ell},G_{y^\prime}} \tilde{B}_{y_{\ell}}+\kappa_{y_{-\ell},G_{y^\prime}} \tilde{B}_{y_{-\ell}}\big] \;, \\
 \frac{\partial \tilde{B}_{x_{\ell}}}{\partial z} =&- j \big[\kappa_{G_{x^\prime},x_{\ell}} \tilde{A}_{G_{x^\prime}}+\kappa_{G_{y^\prime},x_{\ell}} \tilde{A}_{G_{y^\prime}}+\Delta\beta_{x_{\ell}}\tilde{B}_{x_{\ell}}+h^{(b)}_{x_{-\ell},x_{\ell}} \tilde{B}_{x_{-\ell}} +h^{(b)}_{y_{\ell},x_{\ell}} \tilde{B}_{y_{\ell}}+h^{(b)}_{y_{-\ell},x_{\ell}} \tilde{B}_{y_{-\ell}}\big] \;,\\
  \frac{\partial \tilde{B}_{x_{-\ell}}}{\partial z} =&- j \big[\kappa_{G_{x^\prime},x_{-\ell}} \tilde{A}_{G_{x^\prime}}+\kappa_{G_{y^\prime},x_{-\ell}} \tilde{A}_{G_{y^\prime}}+h^{(b)}_{x_{\ell},x_{-\ell}} \tilde{B}_{x_{\ell}}+\Delta\beta_{x_{-\ell}}\tilde{B}_{x_{-\ell}} +h^{(b)}_{y_{\ell},x_{-\ell}} \tilde{B}_{y_{\ell}}+h^{(b)}_{y_{-\ell},x_{-\ell}} \tilde{B}_{y_{-\ell}}\big] \;,\\
   \frac{\partial \tilde{B}_{y_{\ell}}}{\partial z} =&- j \big[\kappa_{G_{x^\prime},y_{\ell}} \tilde{A}_{G_{x^\prime}}+\kappa_{G_{y^\prime},y_{\ell}} \tilde{A}_{G_{y^\prime}}+h^{(b)}_{x_{\ell},y_{\ell}} \tilde{B}_{x_{\ell}} +h^{(b)}_{x_{-\ell},y_{\ell}} \tilde{B}_{x_{-\ell}}+\Delta\beta_{y_{\ell}} \tilde{B}_{y_{\ell}}+h^{(b)}_{y_{-\ell},y_{\ell}} \tilde{B}_{y_{-\ell}}\big] \;,\\
      \frac{\partial \tilde{B}_{y_{-\ell}}}{\partial z} =&- j \big[\kappa_{G_{x^\prime},y_{-\ell}} \tilde{A}_{G_{x^\prime}}+\kappa_{G_{y^\prime},y_{-\ell}} \tilde{A}_{G_{y^\prime}}+h^{(b)}_{x_{\ell},y_{-\ell}} \tilde{B}_{x_{\ell}} +h^{(b)}_{x_{-\ell},y_{-\ell}} \tilde{B}_{x_{-\ell}}+h^{(b)}_{y_{\ell},y_{-\ell}} \tilde{B}_{y_{\ell}}+\Delta\beta_{y_{-\ell}} B_{y_{-\ell}}\big] \;.
 \end{align}
\end{subequations}
Obviously, the Eqs.~\ref{eq:Sixmodes} describe the evolution of the six coupled modes. The mode evolution is dependent on the the couplings and the  modified mode propagation constants, which include the phase shifts. 

The Eqs.~\ref{eq:Sixmodes} can be presented more succinctly in a matrix below
\begin{equation}\label{eq:matrix} 
\begin{aligned} 
{\frac{\partial}{\partial z}\begin{bmatrix}\tilde{A}_{G_{x^\prime}}\\ \tilde{A}_{G_{y^\prime}} \\\tilde{B}_{x_{\ell}}\\ \tilde{B}_{x_{-\ell}}\\ \tilde{B}_{y_{\ell}}\\ \tilde{B}_{y_{-\ell}}\end{bmatrix}}&=- j \begin{bmatrix}\Delta\beta_{G_{x^\prime}} & h^{(a)}_{G_{y^\prime},G_{x^\prime}} & \kappa_{x_{\ell},G_{x^\prime}} & \kappa_{x_{-\ell},G_{x^\prime}} & \kappa_{y_{\ell},G_{x^\prime}} & \kappa_{y_{-\ell},G_{x^\prime}} \\ h^{(a)}_{G_{x^\prime},G_{y^\prime}} & \Delta\beta_{G_{y^\prime}}& \kappa_{x_{\ell},G_{y^\prime}} & \kappa_{x_{-\ell},G_{y^\prime}} & \kappa_{y_{\ell},G_{y^\prime}} & \kappa_{y_{-\ell},G_{y^\prime}} \\ \kappa_{G_{x^\prime},x_{\ell}} & \kappa_{G_{y^\prime},x_{\ell}} & \Delta\beta_{x_{\ell}}& h^{(b)}_{x_{-\ell},x_{\ell}} & h^{(b)}_{y_{\ell},x_{\ell}} & h^{(b)}_{y_{-\ell},x_{\ell}}\\ \kappa_{G_{x^\prime},x_{-\ell}} & \kappa_{G_{y^\prime},x_{-\ell}} & h^{(b)}_{x_{\ell},x_{-\ell}} & \Delta\beta_{x_{-\ell}} & h^{(b)}_{y_{\ell},x_{-\ell}} & h^{(b)}_{y_{-\ell},x_{-\ell}}\\ \kappa_{G_{x^\prime},y_{\ell}} & \kappa_{G_{y^\prime},y_{\ell}} & h^{(b)}_{x_{\ell},y_{\ell}} & h^{(b)}_{x_{-\ell},y_{\ell}} & \Delta\beta_{y_{\ell}} & h^{(b)}_{y_{-\ell},y_{\ell}} \\ \kappa_{G_{x^\prime},y_{-\ell}} & \kappa_{G_{y^\prime},y_{-\ell}} & h^{(b)}_{x_{\ell},y_{-\ell}} & h^{(b)}_{x_{-\ell},y_{-\ell}} & h^{(b)}_{y_{\ell},y_{-\ell}} & \Delta\beta_{y_{-\ell}}\end{bmatrix} 
\begin{bmatrix}\tilde{A}_{G_{x^\prime}}\\ \tilde{A}_{G_{y^\prime}} \\\tilde{B}_{x_{\ell}}\\ \tilde{B}_{x_{-\ell}}\\ \tilde{B}_{y_{\ell}}\\ \tilde{B}_{y_{-\ell}}\end{bmatrix} \;.
\end{aligned} 
\end{equation}
Next, we perform Laplace transform on the above Eq.~\ref{eq:matrix}. For the convenience of expression, we first define a matrix $\Gamma(s)$.
\begin{small}
\begin{equation}\label{eq:Lapmatrix} 
\begin{aligned} 
\Gamma(s)=&\begin{bmatrix}s\!+\!j\Delta\beta_{G_{x^\prime}} & jh^{(a)}_{G_{y^\prime},G_{x^\prime}} & j\kappa_{x_{\ell},G_{x^\prime}} & j\kappa_{x_{\!-\!\ell},G_{x^\prime}} & j\kappa_{y_{\ell},G_{x^\prime}} & j\kappa_{y_{\!-\!\ell},G_{x^\prime}} \\ jh^{(a)}_{G_{x^\prime},G_{y^\prime}} & s\!+\!j\Delta\beta_{G_{y^\prime}}& j\kappa_{x_{\ell},G_{y^\prime}} & j\kappa_{x_{\!-\!\ell},G_{y^\prime}} & j\kappa_{y_{\ell},G_{y^\prime}} & j\kappa_{y_{\!-\!\ell},G_{y^\prime}} \\ j\kappa_{G_{x^\prime},x_{\ell}} & j\kappa_{G_{y^\prime},x_{\ell}} & s\!+\!j\Delta\beta_{x_{\ell}} & jh^{(b)}_{x_{\!-\!\ell},x_{\ell}} & jh^{(b)}_{y_{\ell},x_{\ell}} & jh^{(b)}_{y_{\!-\!\ell},x_{\ell}}\\ j\kappa_{G_{x^\prime},x_{\!-\!\ell}} & j\kappa_{G_{y^\prime},x_{\!-\!\ell}} & jh^{(b)}_{x_{\ell},x_{\!-\!\ell}} & s\!+\!j\Delta\beta_{x_{\!-\!\ell}} & jh^{(b)}_{y_{\ell},x_{\!-\!\ell}} & jh^{(b)}_{y_{\!-\!\ell},x_{\!-\!\ell}}\\ j\kappa_{G_{x^\prime},y_{\ell}} & j\kappa_{G_{y^\prime},y_{\ell}} & jh^{(b)}_{x_{\ell},y_{\ell}} & jh^{(b)}_{x_{\!-\!\ell},y_{\ell}} & s\!+\!j\Delta\beta_{y_{\ell}} & jh^{(b)}_{y_{\!-\!\ell},y_{\ell}} \\ j\kappa_{G_{x^\prime},y_{\!-\!\ell}} & j\kappa_{G_{y^\prime},y_{\!-\!\ell}} & jh^{(b)}_{x_{\ell},y_{\!-\!\ell}} & jh^{(b)}_{x_{\!-\!\ell},y_{\!-\!\ell}} & jh^{(b)}_{y_{\ell},y_{\!-\!\ell}} & s\!+\!j\Delta\beta_{y_{\!-\!\ell}}
\end{bmatrix} \;.
\end{aligned} 
\end{equation}
\end{small}
Then, we get the following equation
\begin{equation}\label{eq:Laplace}
\begin{bmatrix}F_{G_{x^\prime}}(s)\\ F_{G_{y^\prime}}(s) \\ F_{x_{\ell}}(s) \\ F_{x_{\!-\!\ell}}(s) \\ F_{y_{\ell}}(s) \\ F_{y_{\!-\!\ell}}(s)
\end{bmatrix} 
\!=\!\Gamma^{-1}(s)\begin{bmatrix}\tilde{A}_{G_{x^\prime}}(0)\\ \tilde{A}_{G_{y^\prime}}(0) \\ \tilde{B}_{x_{\ell}}(0) \\ \tilde{B}_{x_{-\ell}}(0) \\ \tilde{B}_{y_{\ell}}(0) \\ \tilde{B}_{y_{-\ell}}(0)\end{bmatrix} \;,
\end{equation}
where $F$ means Laplace transform. The inverse matrix $\Gamma^{-1}(s)=\Gamma^{*}(s)/\vert \Gamma(s) \vert$, where $\Gamma^{*}$ means the adjoint matrix and $\vert \Gamma(s) \vert$ means determinant of a matrix.

If we further apply the inverse Laplace transform $L^{-1}$ to the above equation Eq.~\ref{eq:Laplace}, then we can obtain 
\begin{equation}\label{eq:InvLaplace} 
\begin{aligned} 
{\begin{bmatrix}\tilde{A}_{G_{x^\prime}}(z)\\ \tilde{A}_{G_{y^\prime}}(z) \\\tilde{B}_{x_{\ell}}(z)\\ \tilde{B}_{x_{-\ell}}(z) \\ \tilde{B}_{y_{\ell}}(z) \\ \tilde{B}_{y_{-\ell}}(z) \end{bmatrix}}&=
M(z)\begin{bmatrix}\tilde{A}_{G_{x^\prime}}(0)\\ \tilde{A}_{G_{y^\prime}}(0) \\ \tilde{B}_{x_{\ell}}(0) \\ \tilde{B}_{x_{-\ell}}(0) \\ \tilde{B}_{y_{\ell}}(0) \\ \tilde{B}_{y_{-\ell}}(0)\end{bmatrix} \;,
\end{aligned} 
\end{equation}
with
 \begin{equation}
 M(z)=L^{-1}(\Gamma^{-1}(s)) \;.
 \end{equation}
The evolution matrix $M(z)$ describe the filed evolution of the system. Once the initial field distribution is known, the final field distribution of the system can be given.

We continue to expand the equation Eq.~\ref{eq:InvLaplace} into the form of six modes:
\begin{subequations}\label{eq:InvLaplacesix}
\begin{align}
\tilde{A}_{G_{x^\prime}}(z)=\sum_{m} M_{1,m}(z)\tilde{A}_{m}(0) +\sum_{n} M_{1,n}(z)\tilde{B}_{n}(0)\;, \\
\tilde{A}_{G_{y^\prime}}(z)=\sum_{m} M_{2,m}(z)\tilde{A}_{m}(0)+\sum_{n} M_{2,n}(z)\tilde{B}_{n}(0) \;, \\
\tilde{B}_{x_{\ell}}(z)=\sum_{m} M_{3,m}(z)\tilde{A}_{m}(0)+\sum_{n} M_{3,n}(z)\tilde{B}_{n}(0) \;, \\
\tilde{B}_{x_{-\ell}}(z)=\sum_{m} M_{4,m}(z)\tilde{A}_{m}(0) +\sum_{n} M_{4,n}(z)\tilde{B}_{n}(0)\;, \\
\tilde{B}_{y_{\ell}}(z)=\sum_{m} M_{5,m}(z)\tilde{A}_{m}(0)+\sum_{n} M_{5,n}(z)\tilde{B}_{n}(0) \;, \\
\tilde{B}_{y_{-\ell}}(z)=\sum_{m} M_{6,m}(z)\tilde{A}_{m}(0) +\sum_{n} M_{6,n}(z)\tilde{B}_{n}(0)\;,
\end{align}
\end{subequations}
when $M_{i,m}$ ($M_{i,n}$) represents the matrix element corresponding to the mode in the waveguide a (b) in the $i$ row. Each mode amplitude is related to the coupling coefficients between different modes, the modified propagation constant shifts and the initial filed. We can solve the above equations and obtain each mode amplitude, and then calculate field distribution of the system.

The evolution matrix $M(z)$ is dependent on the structure of the two waveguides. When we consider the mode evolution in a practice chip in the next section, it needs to be modified according to the structure.

\section{The Light Field Evolution in the OAM Waveguide}

In order to analyze the evolution of the field in the OAM waveguide in detail, we can divide the asymmetric directional coupler into three segments, as is shown in Fig. ~\ref{Figure S3}. When $0\le z\le L_{1}$, there is no coupling between the two waveguides. However, due to the modified propagation constant shift and coupling between the eigenmodes of the waveguide a, the incident field changes with $z$ during propagation. We can describe the evolution as the following equation according to Eq.~\ref {eq:GCME}
\begin{equation} \label{eq:GCME1}
     \frac{\partial \tilde{A}_{m^\prime}}{\partial z} = -j\Delta{\beta_{m^{\prime}}}\tilde{A}_{m^\prime}- j \sum_{m \neq m^\prime} h^{(a)}_{m,m^\prime}  \tilde{A}_m \;.
\end{equation}

We assume the incident Gaussian field can be written as \begin{equation}\label{eq:GaussianIn}
{\bf E}_{in}(z=0)= \tilde{A}_{G_{x^\prime}}(0)\tilde{\bf E}_{G_{x^\prime}}+\tilde{A}_{G_{y^\prime}}(0)\tilde{\bf E}_{G_{y^\prime}} \;.
\end{equation} 
After $L_{1}$-length propagation evolution, the light field evolves as
\begin{equation}\label{eq:GaussianL1}
{\bf E}(z=L_{1})= \sum_m A_{m}(L_{1})\tilde{\bf E}_{m}e^{-j\beta_{m}L_{1}}=\sum_m \tilde{A}_{m}(L_{1})\tilde{\bf E}_{m}e^{-j\bar{\beta}L_{1}} \;,
\end{equation} 
Where $\tilde{A}_{m}(L_{1})$ can be described as 
\begin{equation}\label{eq:InvLaplaceL1} 
\begin{aligned} 
{\begin{bmatrix}\tilde{A}_{G_{x^\prime}}(L_{1}) \\ \tilde{A}_{G_{y^\prime}}(L_{1}) \\ 0 \\ 0 \end{bmatrix}}&=
M_1(L_{1})\begin{bmatrix}\tilde{A}_{G_{x^\prime}}(0)\\ \tilde{A}_{G_{y^\prime}}(0) \\ 0 \\ 0 \end{bmatrix} \;,
\end{aligned} 
\end{equation}
where $M_1$ is the evolution matrix in the region $0\leq z \leq L_1$. We can derive this evolution matrix $M_1(z)=L^{-1}(\Gamma_1^{-1}(s))$ from the general form given in Eq.~\ref{eq:Lapmatrix} for $\Gamma(s)$, by updating the matrix elements according to the sturcture. And the specific form of $\Gamma(s)$ is determined by Eq. \ref{eq:GCME1}.

\renewcommand{\thefigure}{S\arabic{figure}}
\begin{figure*}[htb!]
\centering
\includegraphics[width=0.77\columnwidth]{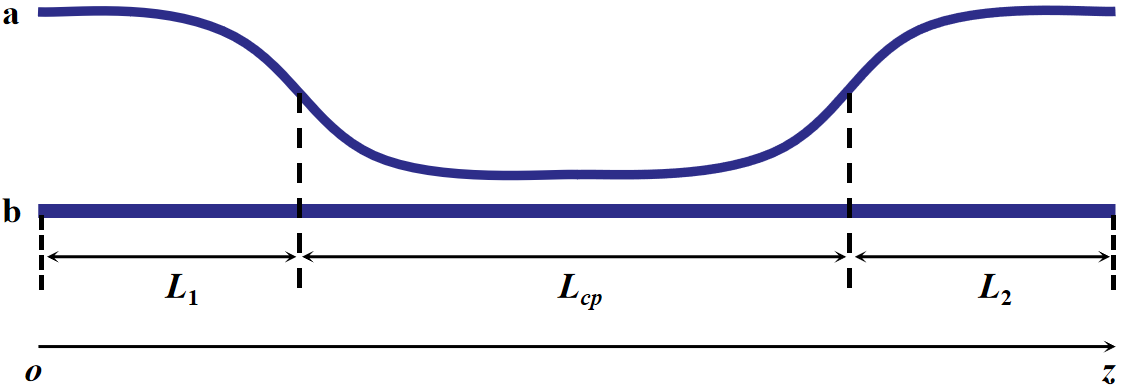}
\caption{\textbf{Schematic of coupling between waveguide a and b}. a (b) is the single-mode (OAM-mode) waveguide. When $0\le z\le L_{1}$ and $L_{1}+L_{cp}\le z \le L$, there is no coupling between the two waveguides ($L$ the length of chip). When $L_{1}\le z \le L_{1}+L_{cp}$, there is a coupling between the two waveguides ($L_{cp}$ the effective coupling length).}
\label{Figure S3}
\end{figure*}

When $L_{1}\le z \le L_{1}+L_{cp}$, there is mutual coupling between the two waveguides. The light field can be written as
\begin{equation}\label{eq:GaussianLcp}
{\bf E}(z=L_{1}+L_{cp})=\sum_n \tilde{B}_{n}(L_{cp})\tilde{\bf E}_{n} e^{-j\bar{\beta}(L_{1}+L_{cp})} \;,
\end{equation} 
where $\tilde{B}_{n}(L_{cp})$ can be expressed as 
\begin{equation}\label{eq:InvLaplaceLcp} 
\begin{aligned} 
{\begin{bmatrix}\tilde{B}_{x_{\ell}}(L_{cp})\\ \tilde{B}_{x_{-\ell}}(L_{cp}) \\ \tilde{B}_{y_{\ell}}(L_{cp}) \\ \tilde{B}_{y_{-\ell}}(L_{cp}) \end{bmatrix}}&=
M_{cp}(L_{cp})\begin{bmatrix}\tilde{A}_{G_{x^\prime}}(L_{1})\\ \tilde{A}_{G_{y^\prime}}(L_{1}) \\ 0 \\ 0\end{bmatrix} \;.
\end{aligned} 
\end{equation}
Here, the evolution matrix $M_{cp}$ can be updated by using Eq.~\ref{eq:Lapmatrix} according to the waveguide structure in the coupling region $L_1 \leq z \leq L_1 +L_{cp}$. It is worth noting that the incident field at this time should be the field after evoluting over $L_1$ in the waveguide a. 

When $L_{1}+L_{cp}\le z \le L$, although there is no coupling between the two waveguides at this time, the field in the OAM waveguide continues to evolve due to the modified propagation constant shift and coupling between the eigenmodes in the waveguide. The field evolution in the OAM waveguide follows the equation according to Eq.~\ref {eq:GCME}
\begin{equation} \label{eq:GCME2}
     \frac{\partial \tilde{B}_{n^\prime}}{\partial z} = -j\Delta{\beta_{n^{\prime}}}\tilde{B}_{n^\prime}- j \sum_{n \neq n^\prime} h^{(b)}_{n,n^\prime} \tilde{B}_n \;.
\end{equation}
The light field in the OAM waveguide can be written as
\begin{equation}\label{eq:GaussianL2}
{\bf E}(z=L)=\sum_n \tilde{B}_{n}(L_{2})\tilde{\bf E}_{n}e^{-j\bar{\beta}L} \;,
\end{equation} 
Where $\tilde{B}_{n}(L_{2})$ can be written as 
\begin{equation}\label{eq:InvLaplaceL2} 
\begin{aligned} 
{\begin{bmatrix} \tilde{B}_{x_{\ell}}(L_{2})\\ \tilde{B}_{x_{-\ell}}(L_{2}) \\ \tilde{B}_{y_{\ell}}(L_{2}) \\ \tilde{B}_{y_{-\ell}}(L_{2}) \end{bmatrix}}&=
M_2(L_{2})\begin{bmatrix} \tilde{B}_{x_{\ell}}(L_{cp})\\ \tilde{B}_{x_{-\ell}}(L_{cp}) \\ \tilde{B}_{y_{\ell}}(L_{cp}) \\ \tilde{B}_{y_{-\ell}}(L_{cp})\end{bmatrix} \;,
\end{aligned} 
\end{equation}
where $M_2$ is the evolution matrix for the region of $L_{cp} + L_1 \leq z \leq L$.

The field at the end of the OAM waveguide can be expressed as ${\bf E}(z=L)=\tilde {\bf E}(z=L) e^{-j\bar{\beta}L}$ and
\begin{equation}\label{eq:GaussianL2}
\tilde {\bf E}(z=L)=\sum_n \tilde{B}_{n}(L_{2})\tilde{\bf E}_{n}=\begin{bmatrix}\tilde{A}_{G_{x^\prime}}(0)\\ \tilde{A}_{G_{y^\prime}}(0) \\ 0 \\ 0 \end{bmatrix}^{T} M^{T}_1(L_{1})M_{cp}^{T}(L_{cp})M_2^{T}(L_{2}) \begin{bmatrix}\tilde{\bf E}_{x_{\ell}} \\ \tilde{\bf E}_{x_{-\ell}} \\ \tilde{\bf E}_{y_{\ell}} \\ \tilde{\bf E}_{y_{-\ell}} \end{bmatrix} \;,
\end{equation} 
$T$ means transpose matrix. If we define 
\begin{equation}\label{eq:GaussianL3}
\begin{bmatrix}\gamma_{x_{\ell}}\\ \gamma_{x_{-\ell}} \\ \gamma_{y_{\ell}} \\ \gamma_{y_{-\ell}} \end{bmatrix}^{T}\!=\!\begin{bmatrix}\tilde{A}_{G_{x^\prime}}(0)\\ \tilde{A}_{G_{y^\prime}}(0) \\ 0 \\ 0 \end{bmatrix}^{T} M_1^{T}(L_{1})M_{cp}^{T}(L_{cp})M_2^{T}(L_{2}) \;.
\end{equation} 
Then, the output field can be further simplified as
\begin{equation}
\tilde{\bf E}(z=L)= \begin{bmatrix}\gamma_{x_{\ell}}\\ \gamma_{x_{-\ell}} \\ \gamma_{y_{\ell}} \\ \gamma_{y_{-\ell}} \end{bmatrix}^{T}\begin{bmatrix}\tilde{\bf E}_{x_{\ell}} \\ \tilde{\bf E}_{x_{-\ell}} \\ \tilde{\bf E}_{y_{\ell}} \\ \tilde{\bf E}_{y_{-\ell}} \end{bmatrix} \;,
\end{equation}
where $\tilde{\mathbf{E}}_{x_{\pm \ell}}={E}_{x_{\pm \ell}}e^{\pm j\ell\phi}{\mathbf e}_{x}$, $\tilde{\mathbf{E}}_{y_{\pm \ell}}={E}_{y_{\pm \ell}}e^{\pm j\ell\phi}{\mathbf e}_{y}$. Normally, $\gamma_{x_{\ell}}$, $\gamma_{x_{-\ell}}$, $\gamma_{y_{\ell}}$, $\gamma_{y_{-\ell}}$ are complex numbers.

We can further rewrite the above equation as:
\begin{equation}\label{eq:OAME_New}
\tilde{\bf E}(z=L) = \big(\gamma_{x_{\ell}} E_{x_{\ell}} e^{j\ell \phi} + \gamma_{x_{-\ell}}E_{x_{-\ell}} e^{-j\ell \phi}\big){\mathbf e}_{x} +\big(\gamma_{y_{\ell}}E_{y_{\ell}} e^{j\ell \phi} +\gamma_{y_{-\ell}}{E}_{y_{-\ell}} e^{-j\ell \phi}\big){\mathbf e}_{y} \;.
\end{equation} 
For convenience, we assume that the optical axis ${\mathbf e}_{x}$ (${\mathbf e}_{y}$) of the waveguide b coincides with the horizontal (vertical) polarization, respectively. Then, the light fields after projecting to horizontal (${\mathbf e}_{x}$), diagonal (${\mathbf e}_{d}=\sqrt{2}({\mathbf e}_{x}+{\mathbf e}_{y})/{2}$), vertical (${\mathbf e}_{y}$) and anti-diagonal (${\mathbf e}_{a}=\sqrt{2}({\mathbf e}_{y}-{\mathbf e}_{x})/{2}$) polarizations can be written as 
\begin{subequations}
\begin{align}
\tilde{\bf E}(z=L) \cdot {\mathbf e}_{x}= & \gamma_{x_{\ell}} E_{x_{\ell}} e^{j\ell \phi} + \gamma_{x_{-\ell}}{E}_{x_{-\ell}} e^{-j\ell \phi} \;,\\
\tilde{\bf E}(z=L) \cdot {\mathbf e}_{d}= & \frac{\sqrt{2}}{2}\big(\gamma_{x_{\ell}} E_{x_{\ell}} e^{j\ell \phi} + \gamma_{x_{-\ell}}{E}_{x_{-\ell}} e^{-j\ell \phi}\big)+ \frac{\sqrt{2}}{2}\big( \gamma_{y_{\ell}}E_{y_{\ell}} e^{j\ell \phi} +\gamma_{y_{-\ell}}{E}_{y_{-\ell}} e^{-j\ell \phi}\big) \;,\\
\tilde{\bf E}(z=L) \cdot {\mathbf e}_{y}= & \gamma_{y_{\ell}}E_{y_{\ell}} e^{j\ell \phi} +\gamma_{y_{-\ell}}{E}_{y_{-\ell}} e^{-j\ell \phi} \;,\\
\tilde{\bf E}(z=L) \cdot {\mathbf e}_{a}= & -\frac{\sqrt{2}}{2}\big( \gamma_{x_{\ell}} E_{x_{\ell}} e^{j\ell \phi} + \gamma_{x_{-\ell}}{E}_{x_{-\ell}} e^{-j\ell \phi})+ \frac{\sqrt{2}}{2}\big( \gamma_{y_{\ell}}E_{y_{\ell}} e^{j\ell \phi} +\gamma_{y_{-\ell}}{E}_{y_{-\ell}} e^{-j\ell \phi}\big) \;.
\end{align}
\end{subequations}

As is shown in Fig. 2 in the Main Text, when the asymmetric direction coupler is excited by right circularly-polarized Gaussian beams by evanescently coupling into the doughnut-shaped waveguide, we obtain a two-lobe intensity distribution. After projecting to horizontal, diagonal, vertical and anti-diagonal polarizations, respectively, it is apparent that this is a scalar light field with diagonal polarization. The two-lobe intensity distributions in near-diagonal direction after projecting to horizontal, vertical polarizations suggest that 
\begin{subequations}
\begin{align}
\gamma_{x_{-\ell}}{E}_{x_{-\ell}}  \simeq & j\gamma_{x_{\ell}}{E}_{x_{\ell}}  \;, \\
\gamma_{y_{-\ell}}{E}_{y_{-\ell}}  \simeq & j\gamma_{y_{\ell}}{E}_{y_{\ell}} \;.
\end{align}
\end{subequations}
It should be noticed that the intensity after anti-diagonal polarization projection is relatively small, in comparison with that projecting to the diagonal polarization. The extinction ratio up to 10.2 dB is observed in the experiment. The field after projecting to the anti-diagonal polarization can be written as
\begin{equation}
\tilde{\bf E}(z=L) \cdot {\mathbf e}_{a}= \frac{\sqrt{2}}{2}\big(\gamma_{y_{\ell}}E_{y_{\ell}}-\gamma_{x_{\ell}} E_{x_{\ell}}\big) \big(e^{j\ell \phi} + je^{-j\ell \phi}\big)\simeq0 \;,
\end{equation}
which means that $\gamma_{y_{\ell}}E_{y_{\ell}}\simeq\gamma_{x_{\ell}} E_{x_{\ell}}$.

With an input of left circularly-polarized Gaussian beam, the generated vortex beam also exhibits a scalar light field with anti-diagonal polarization. Although the two-lobe intensity distribution is uneven, the two lobes near the anti-diagonal direction after respectively projecting to the horizontal, vertical polarizations indicate that 
\begin{subequations}
\begin{align}
\gamma_{x_{-\ell}}{E}_{x_{-\ell}} \simeq & -j\gamma_{x_{\ell}}{E}_{x_{\ell}} \;, \\
\gamma_{y_{-\ell}}{E}_{y_{-\ell}} \simeq & -j\gamma_{y_{\ell}}{E}_{y_{\ell}} \;. 
\end{align}
\end{subequations}
The extinction ratio between the diagonal and anti-diagonal components can reach to $9.6$ dB. The diagonal polarized component is
\begin{equation}
\tilde{\bf E}(z=L) \cdot {\mathbf e}_{d}= \frac{\sqrt{2}}{2}\big(\gamma_{x_{\ell}}{E}_{x_{\ell}}+\gamma_{y_{\ell}}{E}_{y_{\ell}}\big) \big(e^{j\ell \phi} - je^{-j\ell \phi}\big)\simeq0 \;,
\end{equation}
which means that $\gamma_{y_{\ell}}{E}_{y_{\ell}}\simeq -\gamma_{x_{\ell}}{E}_{x_{\ell}}$. 

When the asymmetric direction coupler is excited by horizontally polarized Gaussian beams, we obtian the circularly symmetric first-order vector vortex modes with high quality. The intensity distribution of these modes are annular with a dark core in the center, as shown in Fig. 2. We project the generated vector vortex beam to different polarization with a polarizer. The two-lobe pattern is formed and rotates with the polarizer, showing that the generated beam is a cylindrical vector vortex beam with radial polarization \cite{DornR2003, MaurerR2007, NaidooR2016}. The two-lobe intensity distribution rotates with the polarizer suggests that 
\begin{subequations}
\begin{align}
&\gamma_{x_{-\ell}}{E}_{x_{-\ell}} \simeq \gamma_{x_{\ell}}{E}_{x_{\ell}} \;, \\
&\gamma_{y_{-\ell}}{E}_{y_{-\ell}} \simeq -\gamma_{y_{\ell}}{E}_{y_{\ell}} \;, \\
&\gamma_{y_{\ell}}{E}_{y_{\ell}} \simeq -j\gamma_{x_{\ell}}{E}_{x_{\ell}} \;.
\end{align}
\end{subequations}


\begin{thebibliography}{99}
{
\bibitem{Allen1992}Allen, L. {\it et al.} Orbital angular momentum of light and the transformation of Laguerre-Gaussian laser modes. {\it Phys. Rev. A} {\bf 45}, 8185 (1992).

\bibitem{Wang2012}Wang, J. {\it et al.} Terabit free-space data transmission employing orbital angular momentum multiplexing. {\it Nature Photon.} {\bf 6}, 488-496 (2012).
\bibitem{Bozinovic2013}Bozinovic, N. {\it et al.} Terabit-scale orbital angular momentum mode division multiplexing in fibers. {\it Science} {\bf 340}, 1545-1548 (2013).
\bibitem{Willner2015} Willner, A. E. {\it et al.} Optical communications using orbital angular momentum beams. {\it Advances in Optics and Photonics} {\bf 7}, 66-106 (2015).

\bibitem{Barreiro2008}Barreiro, J. T. {\it et al.} Beating the channel capacity limit for linear photonic superdense coding. {\it Nature Phys.} {\bf 4}, 282-286 (2008).
\bibitem{Dada2011}Dada, A. C. {\it et al.} Experimental high dimensional two-photon entanglement and violations of generalized Bell inequalities. {\it Nature Phys.} {\bf 7}, 677-680 (2011).
\bibitem{Fickler2012}Fickler, R. {\it et al.} Quantum entanglement of high angular momenta. {\it Science} {\bf 338}, 640-643 (2012).
\bibitem{Krenn2014}Krenn, M. {\it et al.} Generation and confirmation of a (100$\times$100)-dimensional entangled quantum system. {\it Proc. Natl. Acad. Sci.} {\bf 111}, 6243-6247 (2014).
\bibitem{Mirhosseini2015}Mirhosseini, M. {\it et al.} High-dimensional quantum cryptography with twisted light. {\it New J. Phys.} {\bf 17}, 033033 (2015).
\bibitem{Bouchard2017}Bouchard, F. {\it et al.} High-dimensional quantum cloning and applications to quantum hacking. {\it Sci. Adv.} {\bf 3}, e1601915 (2017).
\bibitem{Cozzolino2018}Cozzolino, D. {\it et al.} Orbital Angular Momentum States Enabling Fiber-based High-dimensional Quantum Communication. {\it Phys. Rev. Applied} {\bf 11}, 064058 (2019).
\bibitem{D'Ambrosio2012}D'Ambrosio V. {\it et al.} Complete experimental toolbox for alignment-free quantum communication. {\it Nature Commun.} {\bf 3}, 961 (2012).
\bibitem{Abouraddy2006}Abouraddy, A. F., Toussaint, K. C. Jr. Three-dimensional polarization control in microscopy. {\it Phys. Rev. Lett.} {\bf 96}, 153901 (2006).
\bibitem{Roxworthy2010}Roxworthy, B. J., Toussaint, K. C. Jr. Optical trapping with ${\pi}$-phase cylindrical vector beams. {\it New J. Phys.} {\bf 12}, 073012 (2010). 
\bibitem{D'Ambrosio2013}D'Ambrosio V. {\it et al.} Photonic polarization gears for ultrasensitive angular measurements. {\it Nature Commun.} {\bf 4}, 2432 (2013).
\bibitem{Vallone2014} Vallone, G. {\it et al.} Free-space quantum key distribution by rotation-invariant twisted photons. {\it Phys. Rev. Lett.} {\bf 113}, 060503 (2014).
\bibitem{Barreiro2010}Barreiro, J. T. {\it et al.} Remote preparation of single-photon ``hybrid'' entangled and vector-polarization states. {\it Phys. Rev. Lett.} {\bf 105}, 030407 (2010).
\bibitem{Fickler2014}Fickler, R. {\it et al.} Quantum entanglement of complex photon polarization patterns in vector beams. {\it Phys. Rev. A} {\bf 89}, 060301 (2014).
\bibitem{Parigi2015}Parigi, V. {\it et al.} Storage and retrieval of vector beams of light in a multiple-degree-of-freedom quantum memory. {\it Nature Commun.} {\bf 6}, 7706 (2015).
\bibitem{D'Ambrosio2016}D'Ambrosio V. {\it et al.} Entangled vector vortex beams. {\it Phys. Rev. A.} {\bf 94}, 030304(R) (2016).
\bibitem{Cozzolino2019}Cozzolino, D. {\it et al.} Air-core fiber distribution of hybrid vector vortex-polarization entangled states. arXiv:1903.03452 (2019).
\bibitem{Cai2012}Cai, X.-L. {\it et al.} Integrated compact optical vortex beam emitters. {\it Science} {\bf 338}, 363-366 (2012).
\bibitem{Schulz2013}Schulz, S. A. {\it et al.} Integrated multi vector vortex beam generator. {\it Opt. Express} {\bf 21}, 16130-16141 (2013).
\bibitem{Naidoo2016}Naidoo, D. Controlled generation of higher-order Poincar$\acute{e}$ sphere beams from a laser. {\it Nature Photon.} {\bf 10}, 327-332 (2016).
\bibitem{Shao2018}Shao, Z.-K. {\it et al.} On-chip switchable radially and azimuthally polarized vortex beam generation. {\it Opt. Lett.} {\bf 43}, 1263-1266 (2018).
\bibitem{Liu2018}Liu, J. {\it et al.} Direct fiber vector eigenmode multiplexing transmission seeded by integrated optical vortex emitters. {\it Light: Science \& Applications} {\bf 7}, 17148 (2018).
\bibitem{Chen2018}Chen, Y. {\it et al.} Mapping twisted light into and out of a photonic chip. {\it Phys. Rev. Lett.} {\bf 121}, 233602 (2018).
\bibitem{Rafael2008}Rafael, R. G. {\it et al.} Femtosecond laser micromachining in transparent materials. {\it Nature Photon.} {\bf 2}, 219-225 (2008).
\bibitem{Szameit2010}Szameit, A. {\it et al.} Discrete optics in femtosecond-laser-written photonic structures. {\it J. Phys. B: At. Mol. Opt. Phys.} {\bf 43}, 163001 (2010). 
\bibitem{Osellame2012} Osellame, R. {\it et al.} {\it Femtosecond laser micromachining: photonic and microfluidic devices in transparent materials}. (Springer, 2012).
\bibitem{IEEEJQE.22.988}Marcatili, E. Improved coupled-mode equations for dielectric guides. {\it IEEE J. Quantum Electron.} {\bf 22}, 988-993 (1986).
\bibitem{OE.13.1515}Borselli, M. {\it et al.} Beyond the Rayleigh scattering limit in high-Q silicon microdisks: theory and experiment. {\it Opt. Express} {\bf 13}, 1515-1530 (2005).
\bibitem{Suppl-III} See Supplemental Material III for a general coupled mode equations derivation including six eigenmodes.
\bibitem{Huang1994}Huang, W.-P. Coupled-mode theory for optical waveguides: an overview. {\it J. Opt. Soc. Am. A} {\bf 11}, 963-983 (1994).
\bibitem{Okamoto2006}Okamoto, K. Fundamentals of Optical Waveguides (Academic, San Diego, 2006).
\bibitem{PRA.75.023814}Srinivasan, K., Painter, O. Mode coupling and cavity-quantum-dot interactions in a fiber-coupled microdisk cavity. {\it Phys. Rev. A} {\bf 75}, 023814 (2007).
\bibitem{PRA.84.013808}Schmid, S. I. {\it et al.} Pathway interference in a loop array of three coupled microresonators. {\it Phys. Rev. A} {\bf 84}, 013808 (2011).
\bibitem{OE.21.25619}Xia, K.-Y. {\it et al.} Ultrabroadband nonreciprocal transverse energy flow of light in linear passive photonic circuits. {\it Opt. Express} {\bf 21}, 25619-25631 (2013).

\bibitem{Ismaeel2014}Ismaeel, R. {\it et al.} All-fiber fused directional coupler for highly efficient spatial mode conversion. {\it Opt. Express} {\bf 22}, 11610-11619 (2014).
\bibitem{Luo2014}Luo, L.-W. {\it et al.} WDM-compatible mode-division multiplexing
on a silicon chip. {\it Nature Commun.} {\bf 5}, 3069 (2014). 
\bibitem{Wang2014}Wang, J. {\it et al.} Improved 8-channel silicon mode demultiplexer with grating polarizers. {\it Opt. Express} {\bf 22}, 12799-12807 (2014).
\bibitem{Yang2014} Yang, Y.-D. {\it et al.} Silicon nitride three-mode division multiplexing and wavelength-division multiplexing using asymmetrical directional couplers and microring resonators. {\it Opt. Express} {\bf 22}, 22172-22183 (2014).
\bibitem{Mohanty2017}Mohanty, A. {\it et al.} Quantum interference between transverse spatial waveguide modes. {\it Nature Commun.} {\bf 8}, 14010 (2017). 
\bibitem{Suppl-IV} See Supplemental Material IV for analyzing the evolution of the field in the OAM waveguide.
\bibitem{Dorn2003}Dorn, R., Quabis, S., Leuchs, G. Sharper focus for a radially polarized light beam. {\it Phys. Rev. Lett.} {\bf 91}, 233901 (2003).
\bibitem{Maurer2007}Maurer, C. {\it et al.} Tailoring of arbitrary optical vector beams. {\it New J. Phys.} {\bf 9}, 78 (2007). 
\bibitem{Cardano2012}Cardano, F. {\it et al.} Polarization pattern of vector vortex beams generated by q-plates with different topological charges. {\it Appl. Optics} {\bf 51}, C1-C6 (2012).
\bibitem{Diener2018}Diener, R. {\it et al.} Effects of stress on neighboring laser written waveguides in gallium lanthanum sulfide. {\it Appl. Phys. Lett.} {\bf 112}, 111908 (2018).
\bibitem{Tang2019}Tang, H. {\it et al.} Experimental Quantum Stochastic Walks Simulating Associative Memory of Hopfield Neural Networks. {\it Phys. Rev. Applied} {\bf 11}, 024020 (2019).
\bibitem{Suppl-I} See Supplemental Material I for the relationship between the writing laser energy and the propagation constant.
\bibitem{Suppl-II} See Supplemental Material II for the optical axis distribution of the OAM waveguide.
\bibitem{Mehul2016}Mehul, M. {\it et al.} Multi-photon entanglement in high dimensions. {\it Nature Photon.} {\bf 10}, 248-252 (2016).
\bibitem{Erhard2018}Erhard, M. {\it et al.} Twisted photons: new quantum perspectives in high dimensions. {\it Light: Sci. Appl.} {\bf 7}, 17146 (2018). 
\bibitem{D'Ambrosio2017}D'Ambrosio V. {\it et al.} Tunable two-photon quantum interference of structured light. {\it Phys. Rev. Lett.} {\bf 122}, 013601 (2019).
}
\end{thebibliography}

\begin{thebibliography}{99} 
{
\section*{Supplemental References}
\bibitem{OsellameR2012} Osellame, R. {\it et al.} Femtosecond laser micromachining: photonic and microfluidic devices in transparent materials. (Springer, 2012).
\bibitem{RajeshR2010}Rajesh, S., Bellouard, Y. Towards fast femtosecond laser micromachining of fused silica: The effect of deposited energy. {\it Opt. Express} {\bf 18}, 21490-21497 (2010).
\bibitem{LebugleR2015}Lebugle, M. {\it et al.} Experimental observation of NooN state Bloch oscillations. {\it Nature Commun.} {\bf 6}, 9273 (2015).
\bibitem{DienerR2018}Diener, R. {\it et al.} Effects of stress on neighboring laser written waveguides in gallium lanthanum sulfide. {\it Appl. Phys. Lett.} {\bf 112}, 111908 (2018).
\bibitem{TangR2019}Tang, H. {\it et al.} Experimental Quantum Stochastic Walks Simulating Associative Memory of Hopfield Neural Networks. {\it Phys. Rev. Applied} {\bf 11}, 024020 (2019).
\bibitem{ChenY2018}Chen, Y. {\it et al.} Mapping twisted light into and out of a photonic chip. {\it Phys. Rev. Lett.} {\bf 121}, 233602 (2018).
\bibitem{IEEEJQER.22.988}Marcatili, E. Improved coupled-mode equations for dielectric guides. {\it IEEE J. Quantum Electron.} {\bf 22}, 988-993 (1986).
\bibitem{OER.13.1515}Borselli, M. {\it et al.} Beyond the Rayleigh scattering limit in high-Q silicon microdisks: theory and experiment. {\it Opt. Express} {\bf 13}, 1515-1530 (2005).
\bibitem{HuangR1994}Huang, W.-P. Coupled-mode theory for optical waveguides: an overview. {\it J. Opt. Soc. Am. A} {\bf 11}, 963-983 (1994).
\bibitem{OkamotoR2006}Okamoto, K. Fundamentals of Optical Waveguides (Academic, San Diego, 2006).
\bibitem{PRAR.75.023814}Srinivasan, K., Painter, O. Mode coupling and cavity-quantum-dot interactions in a fiber-coupled microdisk cavity. {\it Phys. Rev. A} {\bf 75}, 023814 (2007).
\bibitem{PRAR.84.013808}Schmid, S. I. {\it et al.} Pathway interference in a loop array of three coupled microresonators. {\it Phys. Rev. A} {\bf 84}, 013808 (2011).
\bibitem{OER.21.25619}Xia, K.-Y. {\it et al.} Ultrabroadband nonreciprocal transverse energy flow of light in linear passive photonic circuits. {\it Opt. Express} {\bf 21}, 25619-25631 (2013).
\bibitem{JLTR.5.16}Haus, H. {\it et al.} Coupled-mode theory of optical waveguides. {\it J. Lightwave Technol.} {\bf 5}, 16-23 (1987).
\bibitem{IEEEJQER.23.1689}Weierholt, A. {\it et al.} Eigenmode analysis of symmetric parallel waveguide couplers. {\it IEEE J. Quantum Electron.} {\bf 23}, 1689-1700 (1987).
\bibitem{CorrielliR2014}Corrielli, G. {\it et al.} Rotated waveplates in integrated waveguide optics. {\it Nature Commun.} {\bf 5}, 5249 (2014).
\bibitem{DornR2003}Dorn, R., Quabis, S., Leuchs, G. Sharper focus for a radially polarized light beam. {\it Phys. Rev. Lett.} {\bf 91}, 233901 (2003).
\bibitem{MaurerR2007}Maurer, C. {\it et al.} Tailoring of arbitrary optical vector beams. {\it New J. Phys.} {\bf 9}, 78 (2007).
\bibitem{NaidooR2016}Naidoo, D. Controlled generation of higher-order Poincar$\acute{e}$ sphere beams from a laser. {\it Nature Photon.} {\bf 10}, 327-332 (2016).
}
\end{thebibliography}
\end{document}